\documentclass[
    aps,
    pra,
    twocolumn,
    superscriptaddress,
    showkeys
]{revtex4-2}

\usepackage{tabularray}
\usepackage[colorlinks=true,linkcolor=red,citecolor= blue,urlcolor=blue]{hyperref}
\usepackage{amsmath,graphicx,array,float,amsthm,amssymb}
\usepackage{tikz,color,fullpage,epsf,multirow}
\usepackage{pstricks}
\usepackage{amsmath,graphicx,verbatimbox,array,float}
\usepackage{indentfirst,orcidlink}
\usepackage{braket}
\usepackage{diagbox}
\usepackage{amssymb}
\usepackage{bbold,soul}
\usepackage{comment}

\begin{document}
\title{Optical downlink modeling for low-earth-orbit and medium-earth-orbit satellites under atmospheric turbulence with a quantum-state-tomography\\ use case}

\author{Artur Czerwinski~\!\!\orcidlink{0000-0003-0625-8339}}\email{aczerwin@umk.pl}
\affiliation{Institute of Physics, Faculty of Physics, Astronomy and Informatics, Nicolaus Copernicus University in Torun, ul. Grudziadzka 5, 87-100 Torun, Poland}
\affiliation{STARTOVA UMK Sp. z o.o., ul. Gagarina 7, 87-100 Torun, Poland}
\affiliation{EmpiriQa Sp. z o.o., ul. Gagarina 7/47, 87-100 Torun, Poland}

\author{Jakub J. Borkowski~\!\!\orcidlink{0009-0007-1526-9394}}
\affiliation{Institute of Physics, Faculty of Physics, Astronomy and Informatics, Nicolaus Copernicus University in Torun, ul. Grudziadzka 5, 87-100 Torun, Poland}

\author{Maciej Wiśniewski~\!\!\orcidlink{0009-0008-1262-5186}}
\affiliation{Institute of Physics, Faculty of Physics, Astronomy and Informatics, Nicolaus Copernicus University in Torun, ul. Grudziadzka 5, 87-100 Torun, Poland}

\author{Saeed Haddadi~\!\!\orcidlink{0000-0002-1596-0763}}
\affiliation{School of Particles and Accelerators, Institute for Research in Fundamental Sciences (IPM), P.O. Box 19395-5531, Tehran, Iran}

\begin{abstract}
This paper presents a comprehensive analysis of the link budget for free-space optical systems involving Low Earth Orbit (LEO) and Medium Earth Orbit (MEO) satellites. We develop a detailed model of the satellite-to-ground channel that accounts for the primary physical processes affecting transmittance: atmospheric absorption and scattering, free-space diffraction, and turbulence-induced fluctuations. The study introduces a general method for computing transmittance along a slant path between a satellite and an optical ground station, incorporating zenith angle, slant range, and altitude-dependent attenuation. The proposed framework is intended to support the design and evaluation of space-based optical links and serves as a critical tool for defining technical specifications in satellite communication demonstrators and simulations. Numerical estimates are provided to illustrate the magnitude of losses under typical operational conditions, including the role of aperture averaging. In addition to the link budget analysis, we introduce a satellite-based quantum use case. We propose a scheme for quantum state tomography performed on states generated by an onboard photon source on an LEO or MEO satellite and transmitted to the optical ground station. This approach enables continuous verification of the quality of quantum resources that can be used to perform quantum protocols within quantum information networks.
\end{abstract}


\maketitle

\section{Introduction}
The development of satellite-based free-space optical (FSO) communication systems has become increasingly important in the context of emerging global-scale quantum networks and high-speed classical and quantum data transfer \cite{kaushal2016optical,wehner2018quantum,guiomar2022coherent,parny2023satellite}. Optical communication links offer significant advantages over traditional radio-frequency systems, including broader bandwidth, higher data rates, immunity to electromagnetic interference, and reduced power consumption and beam divergence \cite{hamza2018classification}. However, the performance of such systems is severely influenced by atmospheric effects, particularly turbulence-induced signal fluctuations and photon losses, which complicate reliable signal transmission between satellites and ground stations \cite{andrews2005laser,bohren2008absorption,liang2022link}.

A critical component of system design is the accurate evaluation of the link budget, which quantifies the overall signal attenuation along the optical path. This includes geometric losses due to beam spreading, attenuation due to absorption and scattering in the atmosphere, and additional effects caused by atmospheric turbulence \cite{pirandola2021limits,pirandola2021satellite,Czerwinski2026}. In downlink scenarios, where the signal travels from a satellite to a receiver located at an optical ground station (OGS), turbulence-induced intensity fluctuations can lead to significant temporal and spatial variability in the received power. These fluctuations degrade system performance, especially for quantum communication applications where the detection of single photons is crucial \cite{sharma2019analysis,trinh2022statistical}.

One technique to mitigate the effects of turbulence-induced fluctuations is aperture averaging \cite{fried1967aperture,andrews1992aperture,andrews2000aperture}, where a larger receiver aperture effectively averages out rapid intensity variations caused by small-scale refractive index inhomogeneities \cite{andrews1999theory}. The degree to which aperture averaging can reduce the impact of turbulence depends on several parameters, including the diameter of the telescope, propagation distance, and the turbulence profile along the path \cite{andrews1992aperture,andrews2000aperture,giggenbach2008fading,giggenbach2017scintillation}. Understanding the extent and limitations of this technique is essential for the practical implementation of optical satellite links, particularly when considering different orbital regimes such as Low Earth Orbit (LEO) and Medium Earth Orbit (MEO), which are commonly considered in satellite optical communication studies (see, e.g., Refs.~\cite{bedington2017progress,rashed2020performance,walsh2022demonstration}).

This work presents a detailed numerical study of the influence of aperture averaging on the link budget for optical satellite downlinks. We employ two models to describe turbulence-induced losses: the Intensity Scintillation Index (ISI), representing fluctuations without averaging, and the Power Scintillation Index (PSI), which incorporates the aperture averaging effect following the model by Andrews et al. \cite{andrews1992aperture,andrews2000aperture,andrews2005laser}. Through comparative simulations, we evaluate photon loss as a function of the zenith angle and telescope diameter for both LEO and MEO satellites. In contrast to previous studies that focused on specific configurations, such as assuming a constant telescope diameter \cite{liang2022link} or analyzing a fixed link distance \cite{Naghshvarianjahromi2022} or fixing both aperture size and propagation distance \cite{Kim2022}, our framework accounts for a range of telescope diameters, allowing for a broader and more systematic assessment. In addition, by considering two orbital regimes (LEO and MEO) and evaluating the results as a function of the zenith angle, we explicitly demonstrate how the overall loss depends on propagation distance. This approach highlights the practical relevance of aperture averaging across different orbital altitudes and provides guidance for selecting design parameters that ensure robust and efficient FSO communication systems.

Beyond classical photon loss analysis, we include a quantum use case that demonstrates how the link budget affects the quality of quantum resources distributed from a satellite to OGSs. We propose a satellite-based quantum state tomography (QST) scheme in which an onboard source generates polarization-encoded quantum states that are transmitted to the ground for continuous verification. This approach enables real-time monitoring of state quality during operation. Our contribution lies in connecting the transmittance model with a practical method for assessing quantum state quality under atmospheric loss and turbulence. Such verification may play an important role in future satellite missions that distribute entanglement, single photons, or other quantum resources for quantum communication and sensing.

The structure of the paper is as follows. Section~\ref{theorframework} outlines the theoretical framework used to model atmospheric transmittance in satellite downlinks, including the incorporation of turbulence effects through ISI and PSI models. Section~\ref{results} presents the results of numerical simulations comparing LEO and MEO downlink performance, with a particular focus on the impact of aperture averaging. Section~\ref{usecase} introduces the satellite-based QST scheme, describes the adopted simulation model, and presents numerical results illustrating how the quality of the reconstructed quantum states depends on the properties of the transmitted quantum resources and the communication channel. Two operating scenarios are considered: dark-sky conditions, in which background noise is neglected, and bright-sky conditions, which include background photons originating from sky radiance. Finally, Section~\ref{finalsec} provides concluding remarks and outlines potential directions for future research.

\section{Atmospheric channel modeling for the optical downlink}\label{theorframework}

\subsection{General method for modeling the transmittance}

Consider a satellite orbiting the Earth along a circular trajectory at altitude $H$. From the perspective of an OGS, the zenith angle $\zeta$ is defined as the angle between the local vertical and the line of sight to the satellite. The slant path distance $z$, which represents the optical propagation path from the transmitter to the receiver, depends on both the satellite altitude $H$ and the zenith angle $\zeta$, i.e., $z \equiv z(H, \zeta)$.

The primary objective is to determine the fraction of an optical signal transmitted from the satellite that can be detected by the OGS. The transmittance for a satellite-to-ground optical link is influenced by several independent factors and can be expressed as \cite{Czerwinski2026,trinh2022statistical,Czerwinski2025}
\begin{equation}\label{transmittance1}
    \eta = \eta_{\textrm{int}} \,\eta_{\textrm{atm}} (\zeta) \, \eta_{\textrm{d}} (z)\, I,
\end{equation}
where:
\begin{itemize}
    \item $\eta_{\textrm{int}}$ accounts for the internal losses within the detection system,
    \item $\eta_{\textrm{atm}} (\zeta)$ represents photon losses due to atmospheric absorption and scattering depending on the position of the satellite characterized by the angle zenith $\zeta$,
    \item $\eta_{\textrm{d}} (z)$ corresponds to diffraction-induced losses that depends on the slant distance $z$,
    \item $I$ quantifies turbulence-induced intensity fluctuations.
\end{itemize}

Since the contributions from all four factors are statistically independent, the overall transmittance is given by their multiplicative combination. While $\eta$ expresses the fraction of signal power reaching the detector, it is often convenient to use the logarithmic scale (dB). In this representation, the logarithmic value of transmittance corresponds to photon loss, which is the standard metric in link budget analysis for quantifying received power under given conditions, cf. Ref.~\cite{Czerwinski2025}.

The internal losses, attributed to the inefficiency of the on-ground detection system, are fixed and depend on the implemented technological solutions. For numerical simulations, it is usually assumed that $\eta_{\textrm{int}} = 0.4-0.5$ unless a more precise approximation of this type of loss is known. The remaining three factors, $\eta_{\textrm{atm}} (\zeta)$, $\eta_{\textrm{d}} (z)$, and $I$, stem from actual physical processes, and the formalism for computing their values is explained in the following subsections.

```latex
\subsection{Atmospheric attenuation}

The atmospheric absorption and scattering of an optical beam depend on the composition of the atmosphere and the local atmospheric conditions during optical communication. Absorption is a wavelength-dependent phenomenon, with each wavelength associated with an absorption coefficient determined by the atmospheric chemical composition \cite{giggenbach2022atmospheric}. Therefore, the operating wavelength of an FSO communication system can be selected to minimize this absorption coefficient.

Scattering is typically categorized as Rayleigh or Mie scattering. Rayleigh scattering occurs when light interacts with particles whose radii are much smaller than the optical wavelength. In contrast, Mie scattering arises from interactions with larger particles, such as aerosols, whose radii are comparable to the wavelength of light. Both absorption and scattering contribute to transmittance loss at the selected operating wavelength \cite{andrews2005laser}.

The Beer--Lambert law provides a fundamental model for determining atmospheric transmittance along a propagation path:
\begin{equation}\label{beerlambert}
    \eta_{\textrm{atm}} = e^{- \gamma(h) \, d},
\end{equation}
where $d$ represents the propagation distance, and $\gamma(h)$ is the altitude-dependent attenuation coefficient that accounts for both absorption and scattering \cite{bohren2008absorption}.

The attenuation coefficient $\gamma (h)$ is defined by the absorption and scattering properties of the medium \cite{Fahey2021}:
\begin{equation}
    \gamma (h) = \alpha_m (h) + \beta_m (h) + \alpha_a (h) + \beta_a (h),
\end{equation}
where $\alpha_m(h)$ and $\alpha_a(h)$ denote molecular and aerosol absorption, respectively, while $\beta_m(h)$ and $\beta_a(h)$ denote molecular and aerosol scattering. These coefficients depend on the transmitted wavelength, the relevant cross sections, and the concentrations of the individual particles responsible for absorption or scattering. A common model expresses the attenuation coefficient as \cite{Duntley1948}
\begin{equation}
    \gamma(h) = \alpha_0 \exp(-h/h_0),
\end{equation}
where $h_0$ is the atmospheric scale height, which in the standard atmospheric model takes the value $h_0 = 6600~\mathrm{m}$ \cite{vasylyev2019satellite}. Several sources consider the wavelength $\lambda = 800~\mathrm{nm}$, with the corresponding value $\alpha_0 \approx 5 \times 10^{-6}~\mathrm{m}^{-1}$ representing sea-level conditions \cite{vasylyev2019satellite,pirandola2021limits,pirandola2021satellite}. In this paper, we consider $\lambda = 1550~\mathrm{nm}$, which is a typical wavelength for FSO technologies. This wavelength belongs to a favorable atmospheric transmission window, and Rayleigh scattering is approximately $14$ times weaker than at $\lambda = 800~\mathrm{nm}$. Taking into account the effects of Mie scattering and absorption, we assume $\alpha_0 \approx 10^{-6}~\mathrm{m}^{-1}$ at $\lambda = 1550~\mathrm{nm}$ \cite{Elterman1968,Prokes2010}.

The general law in Eq.~(\ref{beerlambert}) must be modified to describe FSO communication via arbitrary links. First, consider a satellite located exactly at the zenith, so that its slant range $z$ equals its altitude $H$. The transmittance can then be computed as \cite{pirandola2021satellite}
\begin{equation}
\label{zenithett}
\begin{aligned}
\eta_{\textrm{atm}}^{\textrm{zen}}
&= \exp \left( - \int_{0}^{H} \gamma(h) \, d h \right) \\
&\geq \exp \left( -\alpha_0 h_0 \right) \\
&\approx 0.9934.
\end{aligned}
\end{equation}
This provides a lower bound for transmittance applicable to altitudes above which atmospheric density becomes negligible. In practice, for any altitude $H \geq 30~\mathrm{km}$, the atmospheric transmittance for a satellite at the zenith position can be estimated as $\eta_{\textrm{atm}}^{\textrm{zen}} \approx 0.9934$, corresponding to a photon loss of approximately $0.0287~\mathrm{dB}$.

The instantaneous altitude of the signal along the slant path, denoted as $h(z, \zeta)$, is a function of both the slant distance and the zenith angle. Using this definition, the generalized form of the atmospheric transmittance becomes:
\begin{equation}
\begin{aligned}
&\eta_{atm} (H, \zeta)
=\\& \exp \Bigg[
- \alpha_0
\int_0^{z(H, \zeta)} d z' 
\exp \left(
- \frac{h(z', \zeta)}{h_0}
\right)
\Bigg].
\end{aligned}
\end{equation}
where $\alpha_0$ is the sea-level attenuation coefficient and $h_0$ is the characteristic atmospheric scale height.

A widely used approximation for atmospheric transmittance, which simplifies the geometric dependence, is given by \cite{pirandola2021satellite,dequal2021feasibility}:
\begin{equation}\label{finalatmext}
    \eta_{atm} (\zeta) = \left(\eta_{atm}^{zen} \right)^{\sec \zeta},
\end{equation}
where $\eta_{atm}^{zen}$ is the transmittance at zenith ($\zeta = 0$). This expression provides a practical estimation of atmospheric attenuation as a function of the zenith angle $\zeta$, assuming a plane-parallel atmospheric model in which the effective optical thickness increases with the secant of the zenith angle.

\subsection{Diffraction-limited transmittance}

The coefficient $\eta_{\textrm{d}}(z)$ quantifies the transmittance loss due to diffraction-induced beam broadening during free-space propagation \cite{mansuripur2002classical,pirandola2021satellite,pirandola2021limits}. For a Gaussian beam perfectly aligned with the center of a circular receiving aperture of radius $a_R$, the fraction of optical power captured at a distance $z$ is given by \cite{pirandola2021satellite,pirandola2021limits}
\begin{equation}
   \eta_{\textrm{d}}(z) = 1 - \exp \left(-  \frac{2 a_R^2}{w_{\textrm{d}}^2(z)} \right),
\end{equation}
where $w_{\textrm{d}}(z)$ denotes the beam spot size (i.e., the transverse beam radius) after propagating along the distance $z$. This spot size increases due to diffraction and is expressed as
\begin{equation}
    w_{\textrm{d}}(z) = w_0 \sqrt{1 + \left(\frac{z}{z_R}\right)^2},
\end{equation}
where $w_0$ is the beam waist (radius) and $z_R$ is the Rayleigh range defined by
\begin{equation}
    z_R = \frac{\pi w_0^2}{\lambda},
\end{equation}
with $\lambda$ denoting the optical wavelength.

The parameter $w_0$ denotes the effective Gaussian beam waist at the transmitter and should not be interpreted as the physical diameter of the transmitting telescope. In practical optical terminals, the relationship between $w_0$ and the telescope aperture depends on the beam-expansion optics and the degree to which the aperture is illuminated.

This formulation assumes paraxial approximation and describes an ideal, diffraction-limited Gaussian beam. The transmittance $\eta_{\textrm{d}}(z)$ captures the geometric loss arising from the mismatch between the expanded beam spot and the finite-size receiving aperture. As the propagation distance $z$ increases, the beam divergence causes more power to fall outside the receiver aperture, thereby reducing the collected power. This effect is particularly significant in long-distance FSO links such as satellite downlinks.

\subsection{Scintillation effects}

Scintillation can heavily affect an FSO communication link by causing intensity fluctuations in the receiver. Scintillation is caused almost exclusively by small temperature variations in the random medium, resulting in index-of-refraction fluctuations (i.e., optical turbulence).

\subsubsection{Refractive Index Structure Coefficient according to the Hufnagel-Valley (H-V) Model}

The strength of atmospheric turbulence is commonly characterized by the refractive index structure parameter $C^2_n$ (m$^{-2/3}$), which directly influences the severity of scintillation effects. Depending on the value of $C^2_n$, turbulence is typically classified as weak, moderate, or strong. This parameter varies with altitude, geographical location, and time of day. For near-ground horizontal links, $C^2_n$ can often be considered constant, with typical values ranging from $10^{-17}$~m$^{-2/3}$ under weak turbulence to as high as $10^{-13}$~m$^{-2/3}$ in strong turbulence conditions. In contrast, for vertical or slant links, $C^2_n$ generally exhibits a strong altitude dependence \cite{kaushal2016optical}.

For optical carrier wavelengths in the range from 800~nm to 15~$\mu$m, the Hufnagel-Valley (H-V) model is widely used to estimate the altitude-dependent behavior of $C^2_n$. The model is given by the following expression \cite{hufnagel1964modulation,valley1980isoplanatic,calvo2014transmitter,anarthe2023design}:
\begin{equation}
\label{hvmodel}
\begin{aligned}
C_n^2(h)
&= 8.148 \times 10^{-56}\,
v_{\text{rms}}^2\, h^{10} e^{-h/1000} \\
&\quad + 2.7 \times 10^{-16}\, e^{-h/1500} \\
&\quad + C_0\, e^{-H_{\text{OGS}}/700}
e^{(H_{\text{OGS}} - h)/100}.
\end{aligned}
\end{equation}
where $h$ denotes the altitude (height above sea level), $C_0$ is the refractive index structure parameter at sea level (typically taken as $1.7 \times 10^{-14}$~m$^{-2/3}$), $v_{\text{rms}}$ is the root-mean-square wind speed along the optical path, and $H_{\text{OGS}}$ is the altitude above sea level of the OGS.

In many standard references, Eq.~(\ref{hvmodel}) appears with the assumption $H_{\text{OGS}} = 0$, effectively placing the station at sea level \cite{toyoshima2011atmospheric}. However, we adopt a more general form that retains the dependence on $H_{\text{OGS}}$, thereby allowing for scenarios where the ground station or receiver is located at elevated altitudes. This generalization is especially useful when modeling potential attacks in quantum key distribution (QKD) protocols, where an eavesdropper might exploit altitude-dependent channel properties.

The wind speed $v_{\text{rms}}$ can be estimated using the Bufton wind model \cite{bufton1973comparison}. Different values of $v_{\text{rms}}$ are used in the literature to represent various atmospheric conditions. For example, $v_{\text{rms}} = 21$~m/s is often used to model weak-wind scenarios \cite{calvo2014transmitter,Czerwinski2026}. In this paper, we choose $v_{\text{rms}} = 26.25$~m/s, which corresponds to moderate wind conditions. This choice reflects our intention not to restrict the analysis to only the most favorable atmospheric situations.

{In this work, the atmospheric turbulence along the satellite-to-ground path is described by the H–V model. It provides only an average turbulence profile and does not explicitly capture local variations due to weather, time of day, or seasonal effects, which can only be measured experimentally \cite{Osborn2018,Hegde2024}. Nevertheless, the H–V model continues to serve as one of the most established frameworks for FSO link modeling. Its analytical form offers a feasible and tractable basis for numerical simulations, which explains why it is still employed extensively in recent literature, often as a benchmark or reference model. For the purpose of this study, the H–V profile is therefore a justified choice, as it allows us to investigate the link budget under conditions that are broadly representative of typical satellite downlinks. A review of alternative turbulence profiles and their relative merits can be found in Ref. \cite{Quatresooz2025}.

\subsubsection{Intensity Scintillation Index (ISI)}

The fluctuations in optical power, $P$, caused by atmospheric turbulence along an optical propagation path are characterized by the ISI, denoted by $\sigma_I^2$. It is defined as the normalized variance of the optical intensity \cite{andrews2005laser,kaushal2016optical}:
\begin{equation}\label{ISI}
    \sigma_I^2 = \frac{\mathrm{Var} [P(t,p)]}{\left(\mathrm{E}[P(t,p)]\right)^2} = \frac{\langle P^2 \rangle}{ \langle P \rangle^2} -1,
\end{equation}
where $\langle x \rangle$ represents the expected value of $x$  and the arguments $t$ and $p$ represent a specific moment in time and point in space, respectively. The ISI quantifies the relative strength of irradiance fluctuations due to atmospheric turbulence.

The turbulence conditions can be categorized based on the value of the ISI:
\begin{itemize}
    \item $\sigma_I^2 < 1$ -- weak fluctuations,
    \item $\sigma_I^2 \approx 1$ -- moderate fluctuations,
    \item $\sigma_I^2 > 1$ -- strong fluctuations,
    \item $\sigma_I^2 \to \infty$ -- the saturation regime.
\end{itemize}

In the weak-fluctuation regime (where $\sigma_I^2 < 1$), the scintillation index for a horizontal link is directly related to the Rytov variance, given by \cite{andrews2005laser}:
\begin{equation}
    \sigma_I^2 = 1.23 \, C_n^2 \, k^{7/6}\, L^{11/6},
\end{equation}
where $C_n^2$ is the refractive index structure constant, $k = 2 \pi/\lambda$ is the optical wavenumber, and $L$ is the propagation path length. In this regime, the Rytov variance describes the irradiance fluctuations for an unbounded plane wave. For strong-fluctuation conditions, this approach serves as an approximate measure of turbulence strength, increasing with either $C_n^2$, $L$, or both.

For a downlink propagation in the weak-turbulence regime, a more precise theoretical expression for the scintillation index of a plane wave was also derived by Rytov. The Rytov index, $\sigma^2_R$, is computed by integrating $C_n^2 (h)$ from Eq.~(\ref{hvmodel}) along the propagation path \cite{andrews2005laser,giggenbach2008fading,trinh2022statistical}:
\begin{equation}
\label{rytovindex}
\begin{aligned}
\sigma_I^2 \approx \sigma_R^2
&= 2.25\, k^{7/6} \, (\sec \zeta)^{11/6} \\
&\quad \times
\int_{H_{\text{OGS}}}^H
C_n^2 (z)\,
(z - H_{\text{OGS}})^{5/6}
\, d z .
\end{aligned}
\end{equation}
where $\zeta$ is the zenith angle, $H_{\text{OGS}}$ is the same quantity as in Eq.~(\ref{hvmodel}), and $H$ is the transmitter altitude (that is, the satellite altitude), which defines the length of the optical turbulence path.

It is important to note that the weak-fluctuation approximation becomes invalid for large zenith angles and shorter wavelengths. In such cases, where turbulence effects transition to moderate or strong regimes, the scintillation index should be computed using the following empirical formula \cite{andrews2005laser}:
\begin{equation}
\begin{aligned}
\sigma_I^2
&=
\exp \Bigg[
\frac{0.49 \, \sigma_R^2}
{\left(1 + 1.11\, \sigma_R^{12/5} \right)^{7/6}}
\\
&\qquad\quad
+
\frac{0.51 \, \sigma_R^2}
{\left(1 + 0.69\, \sigma_R^{12/5} \right)^{7/6}}
\Bigg]
- 1 .
\end{aligned}
\end{equation}
This expression accounts for nonlinear effects in stronger turbulence conditions, improving accuracy beyond the weak-fluctuation limit.

\subsubsection{Power Scintillation Index (PSI)}

The PSI, $\sigma_P^2$, characterizes the fluctuations in received optical power due to atmospheric turbulence by taking into account the properties of the optical equipment. These power variations can lead to signal fading, affecting the reliability of optical communication links.

It is important to distinguish the PSI from the ISI introduced in the previous subsection. The ISI, $\sigma_I^2$, quantifies fluctuations of the optical intensity at a point-like receiver and therefore describes the intrinsic strength of scintillation generated by atmospheric turbulence. In contrast, the PSI, $\sigma_P^2$, characterizes fluctuations of the total optical power collected by a receiver with a finite aperture. Since a realistic telescope integrates the incident intensity over its receiving surface, rapid spatial fluctuations are partially averaged out. Consequently, the received power typically exhibits smaller fluctuations than the local intensity, leading to $\sigma_P^2 \leq \sigma_I^2$.

The relationship between the PSI, $\sigma_P^2$, and the ISI, $\sigma_I^2$, is quantified by the aperture averaging factor, defined as
\begin{equation}\label{avfactor}
    \mathrm{Av}(D) := \frac{\sigma_P^2}{\sigma_I^2}.
\end{equation}
A value of $\mathrm{Av}(D)=1$ corresponds to the limiting case of a very small aperture, for which aperture averaging is negligible and the PSI coincides with the ISI. As the aperture diameter increases, $\mathrm{Av}(D)$ decreases, indicating that the receiver becomes progressively less sensitive to turbulence-induced intensity fluctuations.

Aperture averaging quantifies the reduction of power fluctuations by increasing the receiver aperture size. A larger aperture effectively integrates the intensity variations over its area, averaging out rapid fluctuations caused by small-scale turbulent eddies. This process helps mitigate the effects of atmospheric turbulence and reduces signal fading \cite{kaushal2016optical}.

To estimate $\mathrm{Av}(D)$, one typically uses models based on the spatial intensity distribution. For the case of a plane wave propagating through weak turbulence, an approximation based on the intensity structure size parameter $\rho_I$ is commonly employed \cite{andrews1992aperture,andrews2000aperture,giggenbach2017scintillation}:
\begin{equation}\label{eqad}
    \mathrm{Av}(D) = \left[1 + 1.062 \, \left( \frac{D}{2\, \rho_I} \right)^2  \right]^{-7/6},
\end{equation}
where $D$ represents the aperture diameter. 

Andrews et al. proposed an approximation in which the intensity structure size parameter $\rho_I$ is taken to be the Fresnel zone size, defined as \cite{andrews2000aperture}
\begin{equation}
    \rho_I = \sqrt{\frac{L}{k}},
\end{equation}
where $L$ is the propagation distance, and $k = 2\pi/\lambda$ is the optical wavenumber corresponding to the wavelength $\lambda$. Substituting this approximation into Eq.~(\ref{eqad}) results in a simplified expression for the aperture averaging factor \cite{andrews1992aperture,andrews2000aperture}:
\begin{equation}\label{eqadAn}
    \mathrm{Av}_A(D) = \left[1 + 1.062 \, \left(\frac{k D^2}{4\, L} \right) \right]^{-7/6}.
\end{equation}
This formulation is particularly useful in numerical simulations, as it provides a direct relationship between the aperture size, wavelength, and propagation distance. It is widely applied in FSO communication systems to optimize receiver design and enhance link performance by mitigating scintillation-induced power fluctuations.

According to Giggenbach et al. \cite{giggenbach2017scintillation}, the intensity structure size parameter $\rho_I$ can be estimated using
\begin{equation}\label{rhoi}
\rho_I = 1.5 \sqrt{\frac{L'}{k}} = 1.5 \sqrt{\frac{\lambda L'}{2 \pi}} \approx 0.6 \sqrt{\lambda L'}.
\end{equation}
In Eq.~(\ref{rhoi}), $\rho_I$ is approximated based on the Fresnel zone size, where $L'$ represents the distance from a dominant turbulent layer rather than the total propagation distance. Unlike Andrews’ approach, which assumes $L$ as the entire path length from the source to the receiver, Giggenbach’s method defines $L'$ as the effective path length from the main turbulent layer—typically the tropopause \cite{giggenbach2017scintillation,andrews2000scintillation}. Alternatively, $\rho_I$ could be determined through a more detailed but uncertain method involving the intensity covariance structure function derived from the $C_n^2$-profile.

Following Eq. (\ref{rhoi}), the aperture averaging factor in Giggenbach’s approach is given by
\begin{equation}\label{averG}
\mathrm{Av}_G(D) = \left[1 + 1.062 \times \frac{k D^2}{9 L'} \right]^{-7/6}.
\end{equation}

The parameter \( L' \) depends on the elevation angle \( \theta \) and is given by \cite{giggenbach2017scintillation}
\begin{equation}\label{issp}
L' = H_d \, \frac{\theta/90^{\circ}}{ (\theta/90^{\circ})^2 + (\theta_{\mathrm{max}}/90^{\circ})^2},
\end{equation}
where \( H_d = 12 \) km represents the height of the tropopause. The factor \( \theta_{\mathrm{max}} = 10^{\circ} \) accounts for the maximum elevation angle at which the turbulence structure size reaches its peak before decreasing. The elevation angle \( \theta \) is related to the zenith angle \( \zeta \) by  
\begin{equation}
\zeta = 90^{\circ} - \theta.
\end{equation}  
Using the elevation angle in Eq.~(\ref{issp}) provides a more intuitive representation of the turbulence layer's influence on aperture averaging.  

As \( \theta \) increases towards \( 90^\circ \) (zenith), the parameter \( L' \) asymptotically approaches \( H_d \), meaning that the dominant turbulence layer is effectively located at the tropopause height. However, for smaller elevation angles (closer to the horizon), \( L' \) increases significantly due to the geometric projection effect, leading to a longer effective turbulence path. The presence of \( \theta_{\mathrm{max}} \) prevents unbounded growth at very low elevations by ensuring a smooth transition between the peak turbulence structure size and its subsequent decline. This correction accounts for the practical behavior of turbulence effects observed in FSO links.

Although Eq.~(\ref{averG}) is a simplified model valid primarily in the weak turbulence regime, neglecting multiple scattering and scintillation saturation effects that become significant at very low elevations—the resulting PSI modeling shows strong agreement with measured values. Notably, this formulation outperforms more general all-regime approximation functions, which are often constrained to horizontal paths and assume a zero inner scale \cite{giggenbach2017scintillation}.

In general, the relationship between the PSI, \( \sigma_P^2 \), and the ISI, approximated by the Rytov index \( \sigma_R^2 \), is nonlinear. A more refined theoretical approach to estimating the aperture averaging factor \( \mathrm{Av} (D) \) in the weak turbulence regime was proposed by Yura et al. This approach accounts for the turbulence distribution along the propagation path and is given by \cite{yura1983aperture,kaushal2016optical}:  
\begin{equation}
    \mathrm{Av}_Y (D) = \left[1 + 1.1\, \left(\frac{D^2}{ \lambda\, h_s\, \sec{\zeta}}  \right)^{7/6}  \right]^{-1},
\end{equation}
where \( h_s \) is the turbulence scale height, defined as  
\begin{equation}
    h_s = \left(\frac{\int_{h_0}^H C^2_n (z) (z - h_0)^2 dz }{\int_{h_0}^H C^2_n (z) (z - h_0)^{5/6} dz} \right)^{6/7}.
\end{equation}
Here, \( h_0 \) represents the reference height, playing the same role as $H_{\text{OGS}}$ in Eq.~(\ref{rytovindex}), and \( C_n^2(z) \) is the refractive index structure parameter as a function of altitude. The term \( \zeta \) represents the zenith angle, ensuring that the slant path effects are properly accounted for.  

Yura’s approach provides a more sophisticated model for aperture averaging by incorporating the altitude-dependent turbulence profile into the estimation of \( A(D) \). This makes it particularly useful for scenarios where turbulence is not uniformly distributed along the path, such as in slant-path links or when atmospheric layering significantly affects wave propagation.

Overall, there are several mathematical approaches to model the value of aperture averaging. Here, we have briefly outlined the mathematical differences between three selected formulas: $\mathrm{Av}_A(D)$, $\mathrm{Av}_G(D)$, and $\mathrm{Av}_Y(D)$. As with many other aspects of atmospheric turbulence theory, there is no single universal model to answer all questions, but rather a variety of methods that approximate reality. When selecting a method for numerical simulations, one must balance reliability and computational efficiency. In this context, the formula provided by Andrews et al., Eq.~(\ref{eqadAn}), aligns optimally with our expectations, as it depends only on the wavelength, propagation distance, and telescope diameter.

\subsubsection{Log-normal probability density function for turbulence modeling}

In this section, we discuss the use of the Log-normal (LN) probability distribution to model the effects of atmospheric turbulence on the intensity of the optical signal. Under the first-order Rytov approximation, the irradiance, $\mathcal{P}$, can be given as \cite{andrews2005laser}:
\begin{equation}\label{irradiance}
    \mathcal{P} = \mathcal{A}^2 e^{2 \chi},
\end{equation}
where \(\mathcal{A}\) is the amplitude of the unperturbed field and $\chi$ denotes the first-order log-amplitude. From Eq.~(\ref{irradiance}), we can represent the log-amplitude as
\begin{equation}
    \chi = \frac{1}{2} \ln \frac{\mathcal{P}}{\mathcal{A}^2}.
\end{equation}

The logarithm of the irradiance is assumed to follow a Gaussian distribution. Consequently, the irradiance itself is modeled as following an LN distribution. This approximation is commonly used to describe intensity fluctuations due to weak-to-moderate atmospheric turbulence. Mathematically, the probability density function (PDF) for irradiance fluctuations can be expressed as a log-normal distribution \cite{andrews2005laser}:

\begin{equation}\label{lognormal1}
    p_{\chi} ( \mathcal{P}) = \frac{1}{2 \sqrt{2 \pi} \,\mathcal{P} \sigma_{\chi}} \exp\left[ - \frac{\left( \ln{\frac{\mathcal{P}}{\mathcal{A}^2}} - 2 \braket{\chi}\right)^2}{8 \sigma_{\chi}^2 } \right],
\end{equation}
where $\mathcal{P}>0$ and \(\sigma_{\chi}^2\) is the variance of the log-amplitude, and \(\braket{\chi}\) represents the mean value of the log-amplitude.

To transform the expression in Eq.~(\ref{lognormal1}) from the irradiance domain to the power domain, we note that the long-term average of the received power $P_0$ is proportional to $\braket{\mathcal{P}}$. Then, from statistics, we know that if a random variable $X$ follows a normal distribution $X \sim \mathcal{N} (\mu, \sigma^2)$, then the random variable $e^{tX}$ has the mean value $\braket{e^{tX}} = \exp (t \mu + 1/2 t^2 \sigma^2)$, which allows us to write a formula for the mean irradiance:
\begin{equation}\label{meanval}
    \braket{\mathcal{P}} = \mathcal{A}^2 \exp \left( 2 \braket{\chi} + 2 \sigma_{\chi}^2  \right).
\end{equation}
Based on Eq.~(\ref{meanval}), one can compute
\begin{equation}
  \ln \frac{P}{P_0}   =  \ln \frac{\mathcal{P}}{\mathcal{A}^2} -  2 \braket{\chi} - 2 \sigma_{\chi}^2,
\end{equation}
which finally allows us to substitute
\begin{equation}\label{equivalence}
    \ln \frac{\mathcal{P}}{\mathcal{A}^2} -  2 \braket{\chi} =  \ln \frac{P}{P_0} + 2 \sigma_{\chi}^2.
\end{equation}

Then, we recognize the relationship between the log-amplitude variance \(\sigma_{\chi}^2\) and the normalized variance of the irradiance, \(\sigma_j^2\). From Ref.~\cite{andrews2005laser}, we have:
\begin{equation}\label{indexes}
    \sigma_j^2 = \exp (4 \sigma_{\chi}^2 ) - 1,
\end{equation}
where \(j\) represents the approach used, with \(j = I\) for the ISI model or \(j = P\) for the PSI model. Rearranging Eq.~(\ref{indexes}), one can express \(\sigma_{\chi}^2\) in terms of \(\sigma_j^2\):

\begin{equation}\label{sigmachi}
   \sigma_{\chi}^2 = \frac{1}{4} \ln (\sigma_{j}^2 + 1).
\end{equation}

Substituting Eqs.~(\ref{equivalence}) and (\ref{sigmachi}) into Eq.~(\ref{lognormal1}), we can re-express the PDF of the received power \(P\) in terms of the PSI or ISI, with respect to the long-term average received power \(P_0\). This results in the following expression for the PDF of the received power \(P\) \cite{giggenbach2017scintillation}:
\begin{equation}
\label{giggenbacheq}
\begin{aligned}
p(P, \sigma_j^2 )
&= \frac{1}{P \sqrt{2 \pi \ln (\sigma_j^2 +1)}} \\
&\times\exp \Bigg[- \frac{\left(\ln\frac{P}{P_0} +\frac{1}{2} \ln (\sigma_j^2+1)
\right)^2}{2 \ln (\sigma_j^2+1)}\Bigg].
\end{aligned}
\end{equation}
The formula \eqref{giggenbacheq} is suitable for modeling power fluctuations in the presence of atmospheric turbulence, as it accounts for turbulence strength and involves scaling of fluctuations based on the aperture size.

In our application, however, we are particularly interested in quantifying the contribution of atmospheric turbulence to the overall transmittance, as denoted in Eq.~(\ref{transmittance1}) by the factor $I$. The quantity $I$ represents a dimensionless random variable describing turbulence-induced intensity fluctuations. Since the deterministic loss mechanisms are already accounted for by the factors $\eta_{\mathrm{int}}$, $\eta_{\mathrm{atm}}$, and $\eta_{\mathrm{d}}$, we require $I$ to describe only fluctuations around the mean transmittance. Therefore, we impose the normalization condition
\begin{equation}
\mathbb{E}[I] = 1,
\end{equation}
which ensures that atmospheric turbulence does not modify the long-term average transmittance of the channel but only introduces stochastic fluctuations around this average value. Note that this condition does not refer to the normalization of the probability density function itself, which is already satisfied by construction. Rather, it specifies that the expectation value of the fluctuation factor $I$ is equal to unity. This approach will allow us to investigate fluctuations of the turbulence-induced component in the overall atmospheric transmittance.

Under the assumption of weak turbulence, we approximate the normalized variance \(\sigma_j^2\) as:
\begin{equation}
       \sigma_{j}^2 = \exp (4 \sigma_{\chi}^2 ) - 1   \approx 4 \sigma_{\chi}^2 \hspace{0.25cm}\Leftrightarrow \hspace{0.25cm} \sigma_{\chi}^2 = \frac{1}{4}  \sigma_{j}^2.
\end{equation}

This approximation is valid in the weak turbulence regime, where the variance of the log-amplitude \(\sigma_{\chi}^2\) is sufficiently small.

Finally, with the assumption of weak turbulence and the corresponding approximations, we can now express the PDF of the relative intensity fluctuations, \(I\), using a log-normal distribution \cite{trinh2022statistical}:

\begin{equation}\label{lognormal}
     p_{\textrm{int}} (I) = \frac{1}{ I \sqrt{2 \pi \sigma^2_j}} \exp\left[ - \frac{\left( \ln I + \sigma^2_j/2 \right)^2}{2 \sigma^2_j} \right],
\end{equation}
where \(\sigma_j^2\) represents the logarithmic variance of intensity fluctuations. The specific choice of \(\sigma_j^2\) depends on whether aperture averaging is taken into account:

\begin{enumerate}
    \item For small apertures, where aperture averaging has little effect, \(\sigma_j^2 = \sigma_I^2\) (ISI-based model with Eq.~(\ref{rytovindex})).
    \item For larger apertures, where aperture averaging reduces the effects of scintillation, \(\sigma_j^2 = \sigma_P^2\) (PSI-based model with Eq.~(\ref{avfactor}) and a specific framework for computing $\mathrm{Av}(D)$).
\end{enumerate}

The distribution Eq.~(\ref{lognormal}) describes the relative intensity fluctuations in terms of a log-normal model, where \(\sigma^2_j\) governs the width of the fluctuation. The transition from the original LN model Eq.~(\ref{lognormal1}) to the power-based model Eq.~(\ref{giggenbacheq}), followed by the final expression for the relative intensity fluctuations, represents a progression from a general log-normal model of irradiance fluctuations to one tailored for weak turbulence and aperture averaging. The relationship between these formulas provides a comprehensive framework for understanding the impact of atmospheric turbulence on optical communication systems.

\section{Results of numerical simulations on atmospheric transmittance}\label{results}

 By convention, the trajectory of a satellite is split such that positive and negative angles describe the ascending and descending segments of the pass, respectively. Moreover, $\zeta = 0^\circ$ corresponds to the satellite being directly overhead. In this study, we restrict our analysis to zenith angles in the range from $-80^\circ$ to $+80^\circ$. This limitation is motivated by the fact that optical communication links suffer the most from atmospheric effects, when the satellite is near the horizon. In particular, as the zenith angle approaches $\pm90^\circ$, the optical path through the atmosphere becomes significantly longer, leading to greater turbulence-induced fluctuations and signal degradation due to attenuation. Therefore, it is a common practice in optical satellite communication studies to exclude these extreme angles from link budget calculations.

Although we exclude only $10^\circ$ on either side of the horizon, this reduction has a noticeable impact on the effective communication time. From the observer’s perspective, the satellite’s angular velocity is lowest when it is near the horizon, meaning it spends more time at higher zenith angles. Thus, the decision to omit these regions plays an important role in estimating realistic link availability and performance. Although our main focus is on link budget analysis, in Fig.~\ref{figuretime}, we illustrate the total time a satellite remains visible to an observer during a zenithal pass, along with the corresponding effective time, defined by restricting the zenith angle range to $-80^\circ$ to $+80^\circ$. The effective time represents the actual duration during which a stable optical communication link with an OGS is feasible. The computational method is based on Ref.~\cite{pirandola2021satellite}.

{More specifically, the calculation is based on the apparent angular motion of the satellite relative to the OGS. Assuming a circular orbit, the relative angular velocity can be written as
\begin{equation}
    \omega_r = \sqrt{\frac{G M_E}{(R_E + H)^3}}- \frac{2\pi}{T_E},
\end{equation}
where $G$ denotes the gravitational constant, $M_E$ is the Earth’s mass, $R_E$ is the Earth’s radius, $H$ is the orbital altitude, and $T_E$ is the Earth’s rotation period. Using this quantity, one can determine the time required for the satellite to move between the zenithal position and an arbitrary zenith angle $\theta$ according to
\begin{equation}
    t(\theta,H)=\omega_r^{-1} \arccos \left(\frac{R_E +z(H,\theta)\cos\theta}{R_E + H}\right),
\end{equation}
where $z(H,\theta)$ denotes the slant range between the satellite and the OGS for a given orbital altitude and zenith angle. The total visibility time for a zenithal pass is obtained by considering the full angular range from $-90^\circ$ to $+90^\circ$, whereas the effective communication time is evaluated by restricting the analysis to the operational interval from $-80^\circ$ to $+80^\circ$. Consequently, the effective duration of the optical link is given by
\begin{equation}
    t_{\mathrm{eff}}=2\, t(80^\circ,H).
\end{equation}

This procedure allows us to estimate the fraction of the satellite pass during which atmospheric attenuation and turbulence remain sufficiently low to support stable optical communication, cf. Refs~\cite{pirandola2021satellite,Czerwinski2026}.

\begin{figure}[t]
    \centering
    \includegraphics[width=0.995\columnwidth]{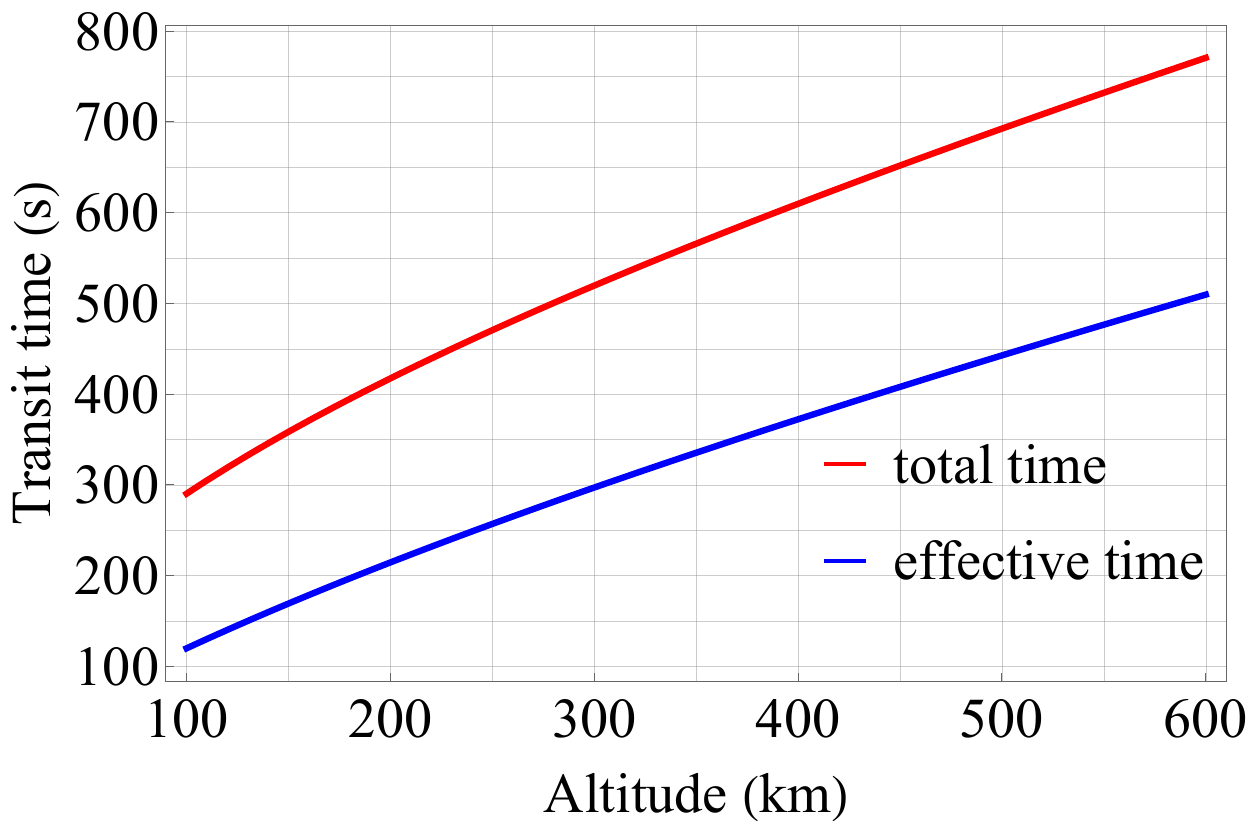}
    \caption{Comparison of total visibility time and effective time for optical communication as a function of satellite altitude. The effective time accounts for zenith angles limited to the range from $-80^\circ$ to $+80^\circ$.}
    \label{figuretime}
\end{figure}

From Fig.~\ref{figuretime}, we observe that for a satellite at an altitude of 500 km the total pass duration is approximately 700 seconds. However, only about 450 seconds are effectively usable for optical transmissions. The remaining 250 seconds need not be wasted; in particular, within the context of quantum protocols, this time can be utilized for data processing or for exchanging classical communication via radio channels. Finally, we note that the angular range used to define effective time can be more restrictive than the one adopted here. For example, in Ref.~\cite{pirandola2021satellite}, only the interval from $-1$~rad to $+1$~rad is considered for quantum communication.

\subsection{Analysis of aperture averaging for LEO and MEO satellites}\label{apavan}

In the following, we shall use the aperture averaging according to the Andrews' model, denoted as $\text{Av}_A(D)$. The effectiveness of aperture averaging depends on both the telescope diameter $D$ and the total propagation distance $L$, as described by the Andrews' model in Eq.~(\ref{eqadAn}). Consequently, for a fixed $D$, the aperture averaging factor can be analyzed as a function of the zenith angle and denoted by $\text{Av}_A(\zeta)$. In this section, we compare aperture averaging for an LEO satellite (altitude of 420 km) and an MEO satellite (altitude of 20 200 km), considering different telescope diameters, as presented in Fig.~\ref{fig:Av_comparison}.

For the LEO case, aperture diameters of $D = 0.2$ m, $0.5$ m, and $1.0$ m were considered, while for the MEO scenario, the diameters were chosen as $D = 0.2$ m, $0.5$ m, and $1.5$ m. The inclusion of a larger aperture in the MEO case ($D = 1.5$ m) was intended to demonstrate the necessary conditions for significant mitigation of intensity fluctuations at greater propagation distances.

\begin{figure}
    \centering
    \includegraphics[width=0.475\textwidth]{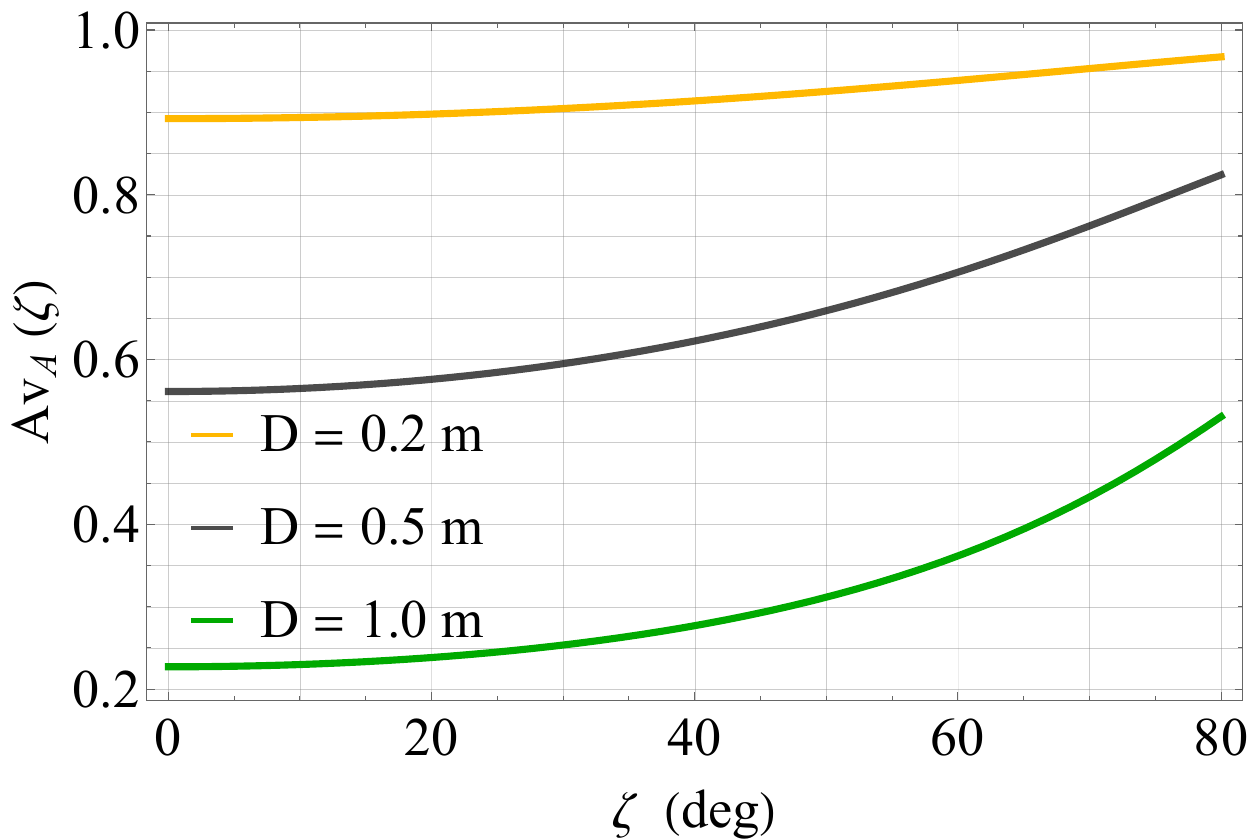}\put(-75,29){(a)~LEO satellite} \quad
    \includegraphics[width=0.475\textwidth]{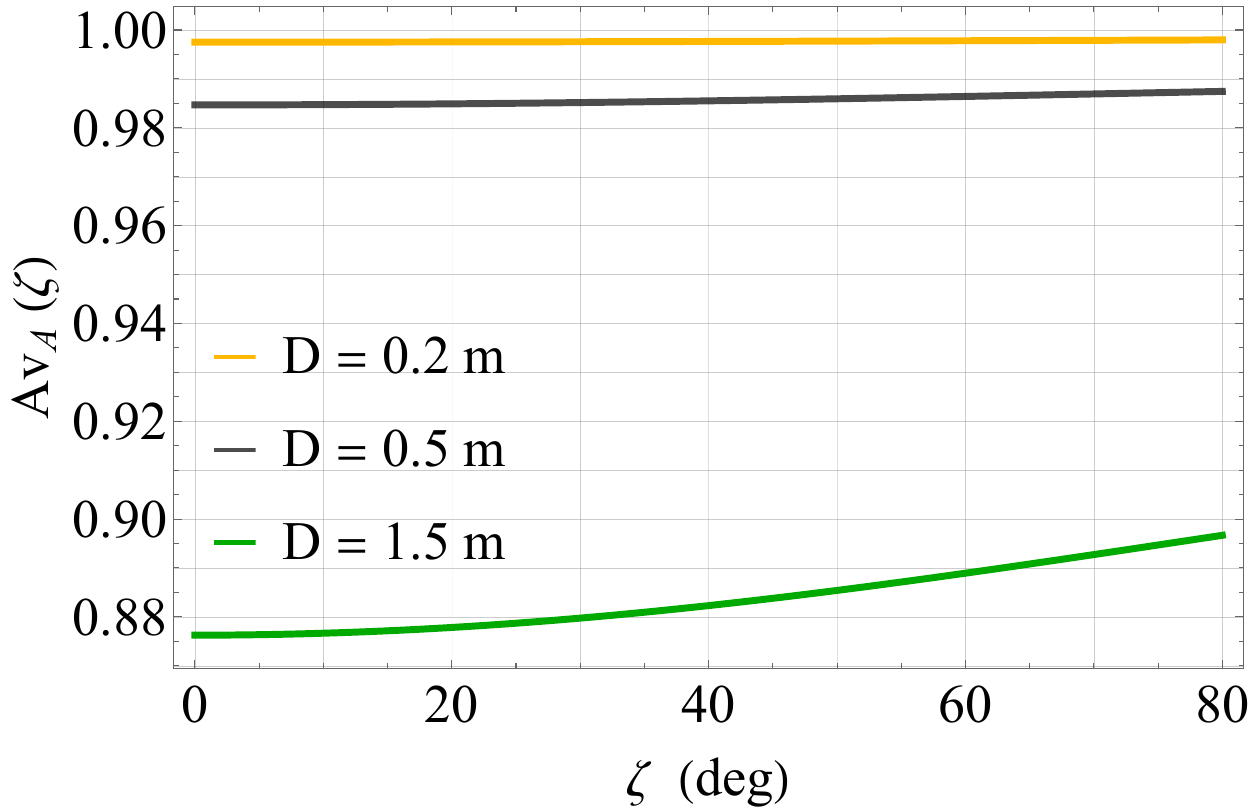}\put(-75,29){(b)~MEO satellite}
    \caption{Aperture averaging according to the Andrews' model \eqref{eqadAn} versus the zenith angle $\zeta$ for (a) an LEO and (b) an MEO satellite, assuming a wavelength of $\lambda = 1550$ nm.}
    \label{fig:Av_comparison}
\end{figure}

The analysis reveals the following key trends:
\begin{itemize}
    \item Stronger aperture averaging for LEO: Due to the shorter propagation distance, even such apertures as $D = 0.2$ m provide significant averaging, reducing the PSI effectively. Increasing the aperture size further enhances this effect, leading to smoother received intensity profiles.
    \item Weaker averaging for MEO: The significantly larger propagation distance in MEO results in a reduced aperture averaging effect for the same aperture sizes, which can be directly explained through the formula Eq. (\ref{eqadAn}). The impact of increasing $D$ is less significant compared to the LEO case.
    \item Effect of increased apertures in MEO: The use of a $D = 1.5$ m telescope in the MEO scenario illustrates that a sufficiently increased aperture can provide noticeable mitigation of intensity fluctuations. This suggests that for optical links at higher altitudes, much larger apertures are required to achieve a comparable level of averaging as in LEO.
\end{itemize}

The observed differences can be explained by the dependency of aperture averaging on the ratio $D^2 / L$. Since $L$ is significantly larger for MEO than for LEO, the same aperture size results in weaker averaging. This necessitates a much larger telescope for MEO to achieve a comparable mitigation effect.

These results highlight an important tradeoff in optical communication system design:
\begin{itemize}
    \item For LEO satellite links, telescope sizes such as $D \approx 0.5$ m are sufficient to suppress intensity fluctuations effectively.
    \item For MEO satellite links, the choice of aperture size becomes more critical, as smaller apertures provide limited mitigation. To achieve significant reduction in fluctuations, telescope diameters of the order of $1.5$ m or larger may be required.
    \item The findings emphasize the necessity of considering aperture size as a key design parameter, particularly for higher-altitude optical links where the benefits of aperture averaging are naturally diminished.
\end{itemize}

\subsection{Link budget in downlink transmission from LEO and MEO satellites: impact of aperture averaging}

For the link budget analysis, we use the same LEO and MEO altitudes as given in Section~\ref{apavan}. The remaining parameters are provided in Table~\ref{tab1}.

\begin{table}[b]
 \caption{Simulation parameters for link budget analysis.}
\def\arraystretch{2.0}
\begin{tabular}{|c|c|c|c|c|c|}
\hline
 \hline   \hspace{0.0cm}$\lambda$ (nm)\hspace{0.0cm} 
& $w_0$ (cm) &$H_{\text{OGS}}$ (m) & $C_0$ (m$^{-2/3}$)& $\bar{v}$ ($\frac{m}{s}$) & $\eta_{\textrm{int}}$ \\\hline
      $1550$  &  $1$ & $65$  & $1.7 \times 10^{-14}$ & $26.25$ & 0.4 \\\hline
\end{tabular}
\label{tab1}
\end{table}

\begin{figure}
    \centering
    \includegraphics[width=0.9995\columnwidth]{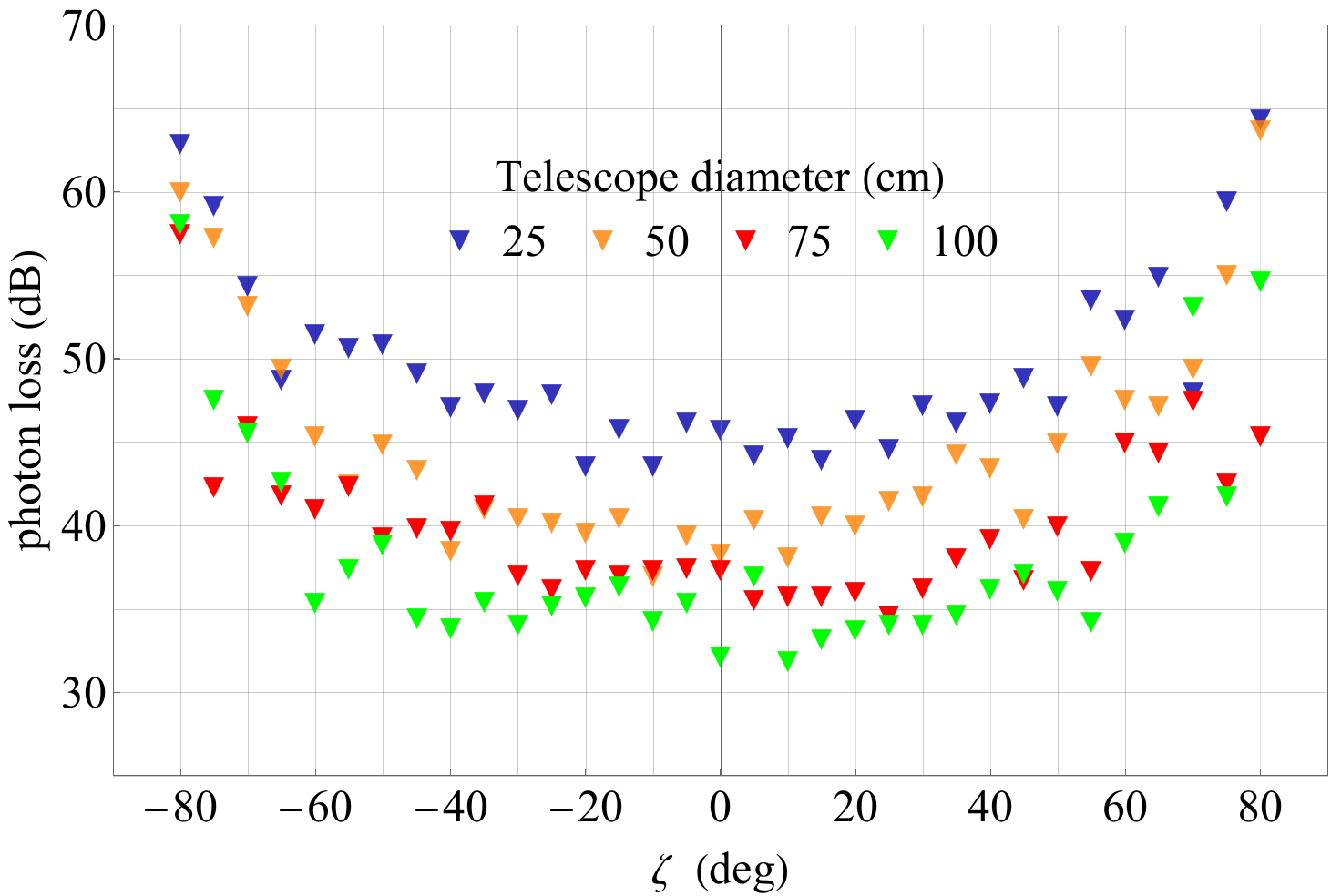}\put(-30,25){(a)~ISI} \quad
    \includegraphics[width=0.9995\columnwidth]{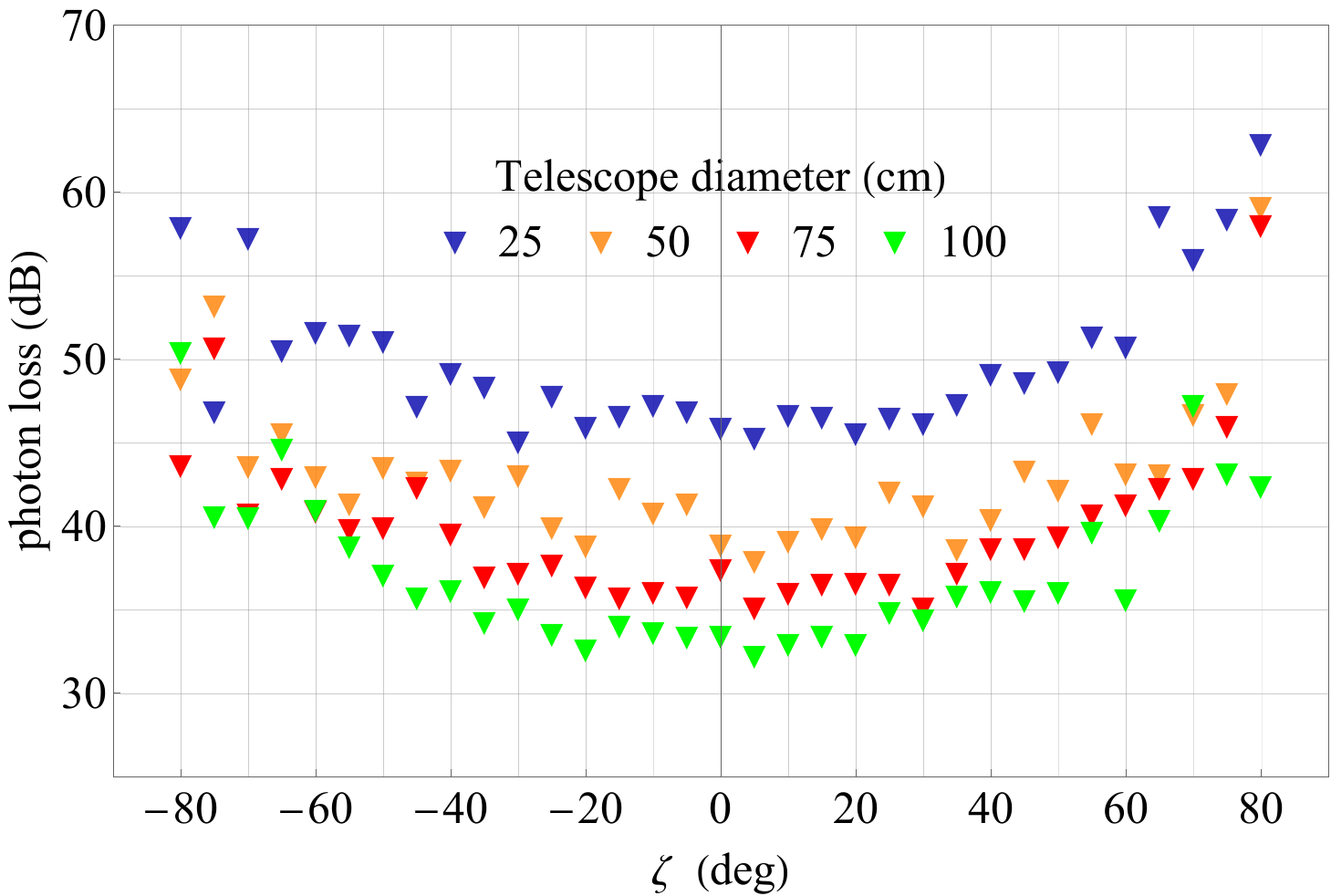}\put(-33,25){(b)~PSI}
    \caption{Photon loss for an LEO satellite pass as a function of the zenith angle $\zeta$, with turbulence-induced fluctuations modeled by (a) ISI and (b) PSI.}
    \label{fig:LEO_losses}
\end{figure}

Figures~\ref{fig:LEO_losses} and \ref{fig:MEO_losses} present the photon loss as a function of the zenith angle $\zeta$ for an LEO and MEO satellite, respectively. We express the link budget in decibels by using a logarithmic scale, which provides a clear comparison of losses across different conditions and telescope sizes. Each figure contains two subplots: (a) modeling turbulence-induced fluctuations using ISI and (b) incorporating aperture averaging via PSI. The small irregular variations visible in the plots originate from the stochastic nature of atmospheric turbulence, as the transmittance factor $I$ is sampled from the corresponding log-normal probability distribution for each simulated zenith angle. The plots cover a range of zenith angles from $-80^\circ$ to $+80^\circ$, excluding the low-elevation region where turbulence effects become excessive. Each subplot shows results for four different telescope diameters: 25 cm, 50 cm, 75 cm, and 100 cm.

In Fig.~\ref{fig:LEO_losses}, the photon loss for the LEO satellite exhibits a characteristic U-shaped dependence on the zenith angle $\zeta$. This behavior arises from the variation in optical path length: near the zenith, the atmospheric path is shortest, resulting in minimal losses, while at larger zenith angles the signal must travel through a significantly longer portion of the atmosphere, increasing attenuation according to Eq.~(\ref{finalatmext}).

When turbulence fluctuations are modeled using ISI alone [Fig.~\ref{fig:LEO_losses}(a)], the losses show higher variability. However, with PSI applied [Fig.~\ref{fig:LEO_losses}(b)], the impact of aperture averaging becomes evident. The curves are noticeably smoother, indicating a reduction in the variance of intensity fluctuations. This effect is most significant for larger telescopes, where aperture averaging is more effective in suppressing scintillation-induced fluctuations.

The difference between ISI and PSI is minor for small apertures (25 cm and 50 cm), as the averaging effect is weak in this regime. However, as the telescope diameter increases to 75 cm and 100 cm, the impact of aperture averaging becomes more apparent, leading to a clear reduction in turbulence-induced variability. This behavior is consistent with the dependence of aperture averaging on aperture size: larger telescopes integrate over more turbulent eddies, effectively smoothing intensity fluctuations. This effect is particularly relevant in LEO downlinks, where the relatively short propagation distance allows aperture averaging to significantly mitigate turbulence-induced power variations. As a result, the received power is closer to the predictions of deterministic formulas, with reduced impact from random fluctuations.

\begin{figure}[t]
    \centering
    \includegraphics[width=0.9995\columnwidth]{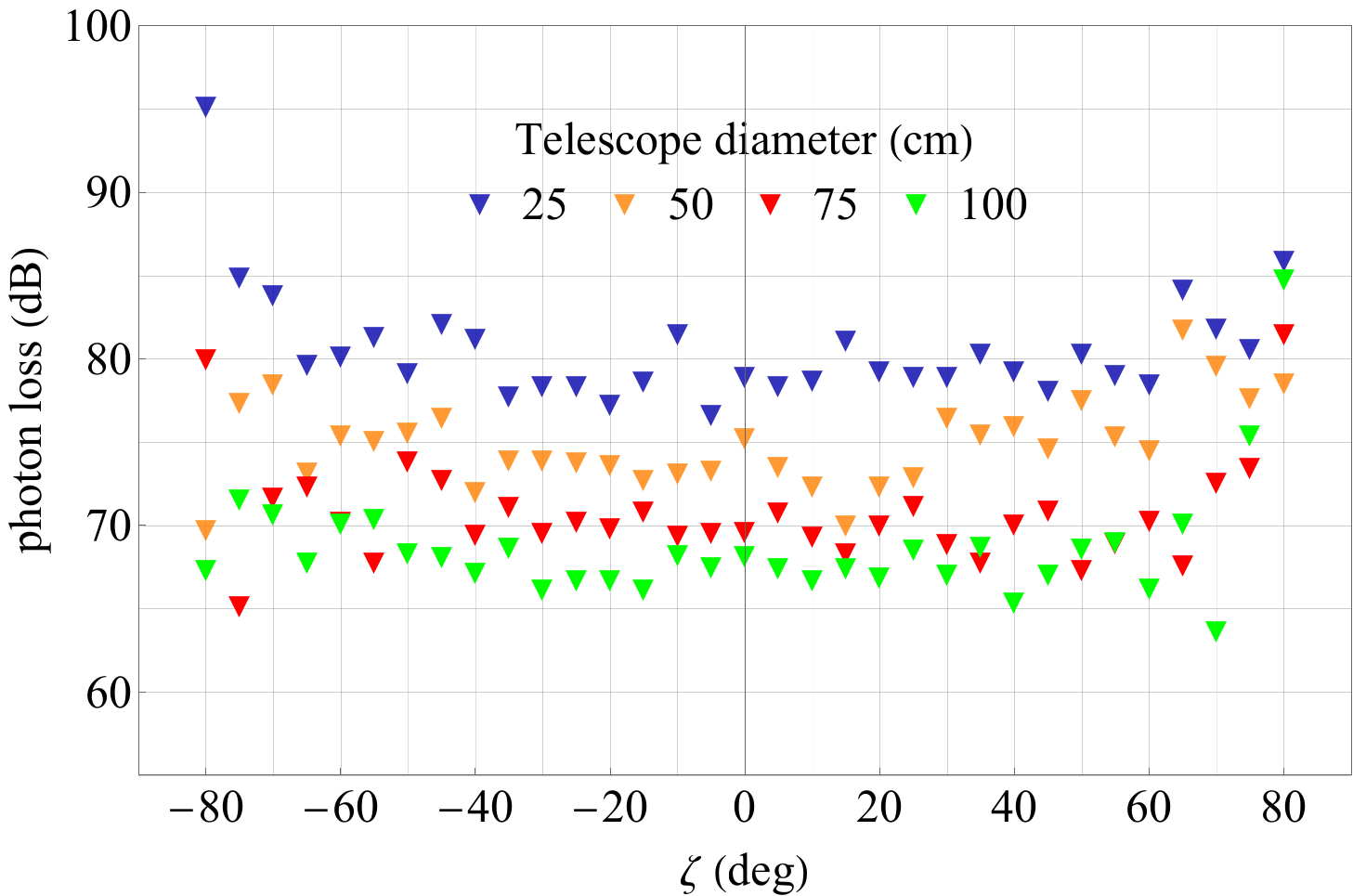}\put(-30,25){(a)~ISI} \quad
    \includegraphics[width=0.9995\columnwidth]{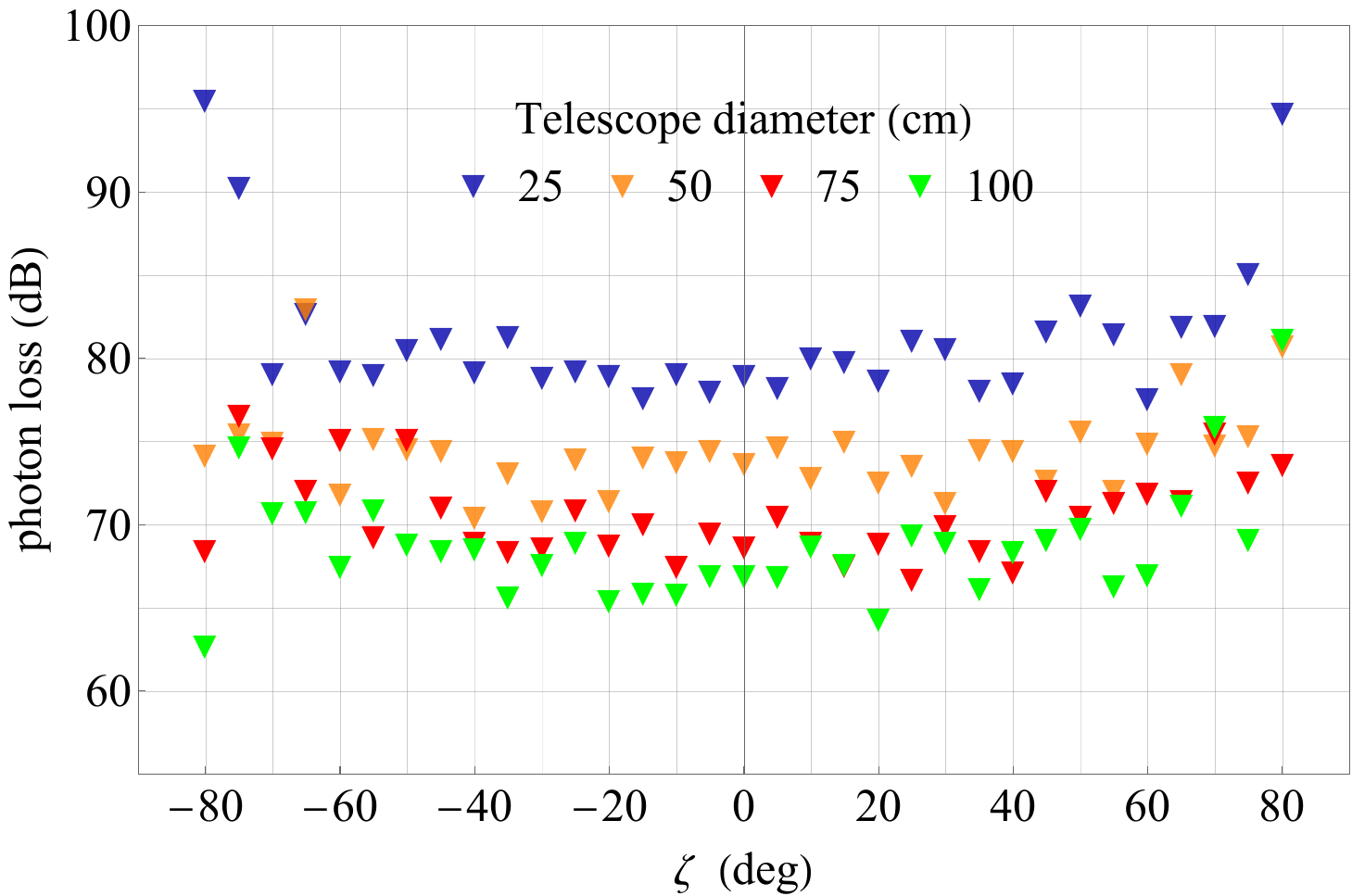}\put(-33,25){(b)~PSI}
    \caption{Photon loss for an MEO satellite pass as a function of the zenith angle $\zeta$, with turbulence-induced fluctuations modeled by (a) ISI and (b) PSI.}
    \label{fig:MEO_losses}
\end{figure}

In contrast, Fig.~\ref{fig:MEO_losses} shows the photon loss for the MEO satellite, where the dependence on the zenith angle is much weaker. This difference stems from the fact that, in MEO, the total optical path is already dominated by free-space propagation outside the atmosphere, making atmospheric path variations relatively less important. As a result, photon loss remains nearly constant across zenith angles, although more noticeable variations in transmittance appear at larger values of $\zeta$. Most importantly, the overall photon losses are significantly higher for MEO-to-Earth transmission compared to LEO. For example, at the satellite's zenith position ($\zeta = 0^\circ$), losses for LEO range from 30–45 dB depending on the telescope diameter, whereas for MEO, the link budget at the same zenith position must assume losses of 65–80 dB.

Furthermore, the difference between ISI and PSI is much less evident, as aperture averaging becomes less effective over long-distance propagation. This outcome is consistent with expectations from aperture averaging theory. Since the propagation path for an MEO satellite is significantly longer, turbulence-induced intensity fluctuations are less affected by averaging. Even for the largest telescope diameter (100 cm), the reduction in variance is marginal. This result reinforces the conclusion that aperture averaging is most useful over shorter propagation distances, such as those in LEO downlinks, but offers limited mitigation of turbulence fluctuations in MEO scenarios.

In both LEO and MEO cases, increasing the telescope diameter consistently reduces photon loss. This is expected, as a larger aperture collects more light, increasing overall signal strength. For MEO satellites, while larger apertures still help reduce losses by improving collection efficiency, they do not substantially suppress turbulence-induced variability.

The analysis of Figs.~\ref{fig:LEO_losses} and \ref{fig:MEO_losses} highlights the role of aperture averaging in satellite downlink transmission. For LEO satellites, aperture averaging significantly smooths the loss curves, demonstrating its usefulness in mitigating turbulence-induced fluctuations. In contrast, for MEO satellites, the effect is much weaker due to the longer propagation distance. These findings suggest that aperture averaging should be considered a key factor in the design of optical links for LEO applications but is of limited relevance for MEO scenarios.

\subsection{Discussion: Link-budget analysis}

In this work, our analysis concentrates on turbulence-induced irradiance fluctuations and their effect on the received power at the aperture. In practical FSO communication links, however, the ultimate figure of merit is often the power coupled into an optical fiber. This process is additionally limited by turbulence-induced phase distortions that degrade the coupling efficiency. A widely used strategy to mitigate these effects is adaptive optics \cite{Vedrenne2016,Canuet2017,Osborn2021}, which can partially restore the wavefront and enhance fiber coupling efficiency. In particular, a recent publication has shown that applying tip/tilt pre-distortion can reduce the mean total uplink losses by 3–5 dB compared to the case without pre-distortion AO \cite{Hristovski2024}. The same study further demonstrated that high-order pre-distortion significantly decreases the PSI by nearly an order of magnitude, in weak to moderate turbulence conditions. While highly relevant, such analyses extend beyond the scope of the present study. We therefore restrict ourselves to modeling scintillation effects at the aperture and leave the integration of adaptive optics into the framework as a goal for future research.

The model proposed in this paper is simulation-based, allowing us to estimate the range of photon loss by including factors enumerated in Eq.~(\ref{transmittance1}). Overall, the obtained photon loss values fall within the general intervals reported in the literature for satellite downlinks, e.g., Ref.~\cite{Ntanos2021}. Direct comparison with experimental data is challenging, as it is difficult to identify missions with parameters identical to those considered here. Nevertheless, available results suggest a consistent trend: experimentally observed losses are generally higher than theoretical predictions.

For example, in Ref.~\cite{Takenaka2017}, the photon loss was theoretically estimated to be approximately $60$~dB for an orbital altitude of $650$~km with a $1$~m ground telescope. However, experimental measurements indicated significantly higher losses, with link budgets ranging from $74.2$~dB to $90.9$~dB. This case illustrates that theoretical estimates should be regarded as lower-bound values, since additional real-world effects contribute to higher overall loss.

These observations confirm that while theoretical modeling provides valuable boundary estimates for photon loss, practical FSO links must account for excess effects. In particular, scintillation effects in urban areas tend to be greater than the theoretical predictions. It can significantly influence achievable key rates in quantum communication protocols. Future system design and key management strategies for QKD should therefore incorporate such uncertainties to ensure robust and reliable operation.

It is also important to address misconceptions in the literature regarding FSO link modeling. For instance, a recent paper~\cite{sengupta2024quantum} models atmospheric loss by treating the FSO channel as an optical fiber with an attenuation coefficient of $0.07$~dB/km and applying only the Beer–Lambert law. This oversimplified approach neglects two key aspects. First, it ignores geometric diffraction losses, which become significant over propagation distances up to 200~km, as emphasized in a recent comment~\cite{czerwinski2024comment}. Second, it disregards turbulence-induced fluctuations. In reality, atmospheric transmittance is governed by stochastic processes such as intensity scintillation and cannot be accurately described by purely deterministic models. We consider the present publication, written in a style close to an academic tutorial, as a tool to help avoid such modeling mistakes.

Finally, the results presented here have implications for intensity modulation/direct detection protocols in optical communications~\cite{ref1,ref2}. In the simplest binary modulation schemes, two intensity levels are used to encode bit values $0$ and $1$. Turbulence-induced intensity fluctuations introduce errors in the encoding/decoding process~\cite{Czerwinski2026}. A better understanding of the aperture averaging effect can help quantify the extent to which it mitigates errors due to scintillation, thereby enabling the design of more robust FSO communication systems.

\section{Satellite-Based QST}\label{usecase}

\subsection{QST as a satellite-based quantum use case}

In order to assess how atmospheric losses and turbulence affect the quality of quantum resources distributed from a satellite, we employ a QST framework as a diagnostic use case. The goal is to reconstruct an unknown quantum state from a set of measurement outcomes and evaluate how accurately this reconstruction can be performed under realistic downlink conditions. As in standard QST protocols, we rely on symmetric informationally complete positive operator-valued measurements (SIC-POVMs) \cite{Renes2004,Fuchus2017}. For qubits, the measurement consists of four operators, while higher-dimensional states or multi-photon systems would require larger operator sets obtained by taking tensor products of the single-photon SIC-POVMs \cite{PaivaSanchez2010}. Let $\{M_k\}_{k=1}^{4}$ denote the complete set of measurement operators. For each tomographic measurement the satellite source emits $\mathcal{N}$ identically prepared photons, each in the unknown state described by a density matrix $\rho_x$.

In the present work, we consider the regime of bulk quantum state tomography, where the reconstruction is performed from measurement statistics collected over an ensemble of identically prepared quantum systems rather than from individual single-photon events. Operationally, this means that for each tomography round, the satellite source emits four optical pulses corresponding to the four SIC-POVM measurement operators, and each pulse contains, on average,} \(\mathcal{N}\) photons prepared in the same quantum state. Consequently, the parameter \(\mathcal{N}\) should be interpreted as the effective ensemble size available for a single tomographic measurement setting \cite{Rehacek2004,Pimenta2013,Sedziak2020}.

The expected photon counts predicted by the Born rule are
\begin{equation}
    e_k = \left\lceil \mathcal{N}\; \mathrm{tr}\!\left( M_k \rho_x \right) \right\rfloor,
\end{equation}
where $\lceil a \rfloor$ denotes rounding to the nearest integer and the density matrix $\rho_x$ is parameterized via a Cholesky decomposition to ensure positivity and unit trace \cite{James2001,Altepeter2005}.

In practice, measurements are affected by statistical fluctuations originating from the discrete nature of photon detection. To emulate realistic detection events, we assume that the measured counts follow independent Poisson distributions, reflecting photon shot noise \cite{Hasinoff2014}. Shot noise is an intrinsic and unavoidable source of uncertainty in photon-counting measurements and constitutes a fundamental limitation of optical detection systems. Even if the optical source emits pulses with a well-defined average photon number, the exact number of detected photons fluctuates randomly from one realization to another. Thereby, the detector cannot determine the photon number with perfect precision. Similar shot-noise-limited models are commonly employed in both quantum and classical optical communication systems, including optical key distribution protocols.

For an input state $\rho_{\mathrm{in}}$, the registered counts are sampled as
\begin{equation}\label{measuredcounts}
    m_k \sim \mathrm{Pois}(n_k), \qquad 
    n_k = \left\lceil \widetilde{\mathcal{N}}\; \mathrm{tr}\!\left( M_k \rho_{\mathrm{in}} \right) \right\rfloor,
\end{equation}
where $\widetilde{\mathcal{N}}$ is the number of photons that successfully reach the ground receiver. In this framework, the measured counts corresponding to different SIC-POVM operators are treated as statistically independent random variables. The Poissonian assumption implies that the variance of the measured counts is equal to the corresponding mean value,
\begin{equation}
\mathrm{Var}(m_k)=n_k,
\end{equation}
which means that the standard deviation (SD) associated with shot noise scales as
\begin{equation}
\sigma_{\mathrm{shot}} \sim \sqrt{\widetilde{\mathcal{N}}},
\end{equation}
whereas the useful signal itself scales linearly with the photon number. As a result, the signal-to-noise ratio improves proportionally to
\begin{equation}
\mathrm{SNR} \sim \sqrt{\widetilde{\mathcal{N}}}.
\end{equation}
Hence, the influence of shot noise becomes particularly significant for low photon budgets, where statistical fluctuations can substantially reduce the fidelity of the reconstructed quantum state.

In the present work, $\widetilde{\mathcal{N}}$ is determined from the atmospheric transmittance predicted by the link budget model, rather than from a simple Beer-Lambert attenuation model used in fiber-based analyses, cf. Ref.~\cite{Czerwinski2022}. Therefore, we substitute $\widetilde{\mathcal{N}} = \eta \mathcal{N}$, where $\eta$ represents the atmospheric transmittance according to Eq.~(\ref{transmittance1}). This substitution allows us to map the performance of QST directly to satellite geometries. In particular, since the atmospheric transmittance depends strongly on the zenith angle $\zeta$, the effective photon number $\widetilde{\mathcal{N}}$ becomes a function of the satellite’s position during a pass. Stronger atmospheric losses and turbulence reduce $\widetilde{\mathcal{N}}$, thereby increasing the relative impact of shot noise on the tomographic reconstruction. Thus, QST enables us to investigate how the ability to reconstruct quantum states varies as a function of $\zeta$.

In our model, we assume that the transmission of four pulses through the atmosphere (one pulse per measurement operator) occurs within the atmospheric coherence time, which is typically on the order of milliseconds \cite{Farley2022,Czerwinski2026}. Consequently, the tomographic procedure for a single input state is carried out under effectively constant atmospheric conditions. In this regime, a single value of the atmospheric transmittance $\eta$ applies to all four measurement operators $\{M_1, M_2, M_3, M_4\}$. 

Given the simulated measurement outcomes $\{m_k\}$, we reconstruct the density matrix by minimizing the least-squares cost function
\begin{equation}
    f_{\mathrm{LS}}(t_1,t_2,\dots)
    = \sum_{k=1}^{4} \left( e_k - m_k \right)^{2},
\end{equation}
where $\{t_j\}$ denotes the set of real parameters defining the Cholesky decomposition of $\rho_x$ \cite{Acharya2019,Czerwinski2021}. To quantify how well the reconstruction approximates the true state, we compute the Uhlmann--Jozsa fidelity \cite{Nielsen2000,Jozsa1994,Uhlmann1976},
\begin{equation}
    F\left[\rho_{\mathrm{in}}, \rho_x\right]
    = \left( \mathrm{tr}\,\sqrt{\sqrt{\rho_{\mathrm{in}}}\,\rho_x\,\sqrt{\rho_{\mathrm{in}}}} \right)^{2}.
\end{equation}

To obtain statistically meaningful performance indicators, we draw a representative ensemble of $220$ input qubits $\rho_{\mathrm{in}}$, simulate QST for each of them at a given zenith angle, and compute the ensemble-average fidelity along with its sample SD. This procedure results in a fidelity curve $F_{\mathrm{av}}(\zeta)$, which quantifies how the efficiency of quantum state reconstruction deteriorates (or recovers) as the satellite moves across the sky during a pass. In this way, QST serves as a practical quantum-level use case for the optical link budget derived in the preceding sections.

\subsection{Results of numerical simulations: dark sky conditions}
\begin{figure*}[t]
    \centering

    \begin{minipage}{0.48\textwidth}
        \centering
        \includegraphics[width=\linewidth]{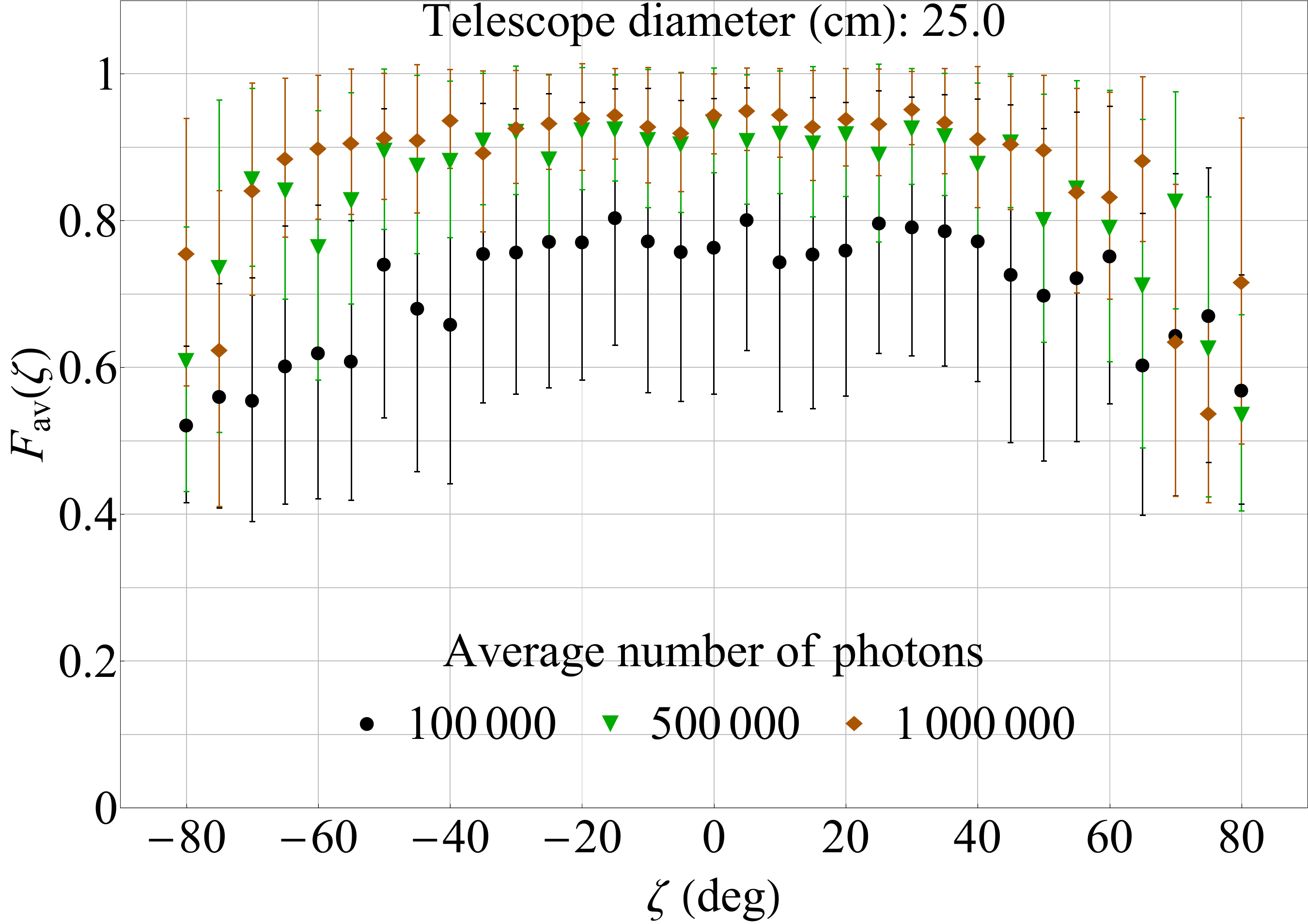}
        \put(-20,26){(a)}\\[2pt]
    \end{minipage}
    \hfill
    \begin{minipage}{0.48\textwidth}
        \centering
        \includegraphics[width=\linewidth]{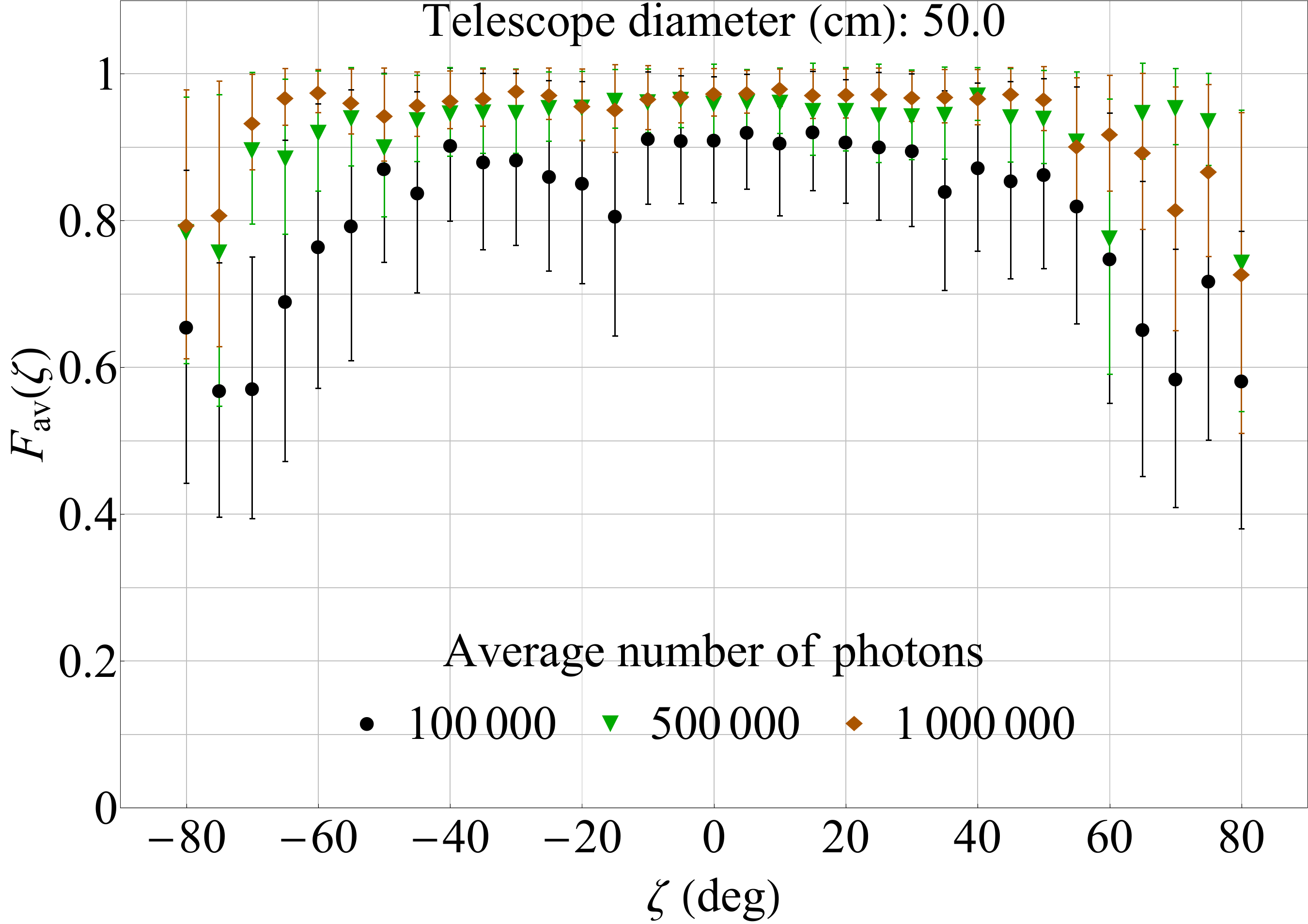}
        \put(-20,26){(b)}\\[2pt]
    \end{minipage}

    \vspace{1.8em}

    \begin{minipage}{0.48\textwidth}
        \centering
        \includegraphics[width=\linewidth]{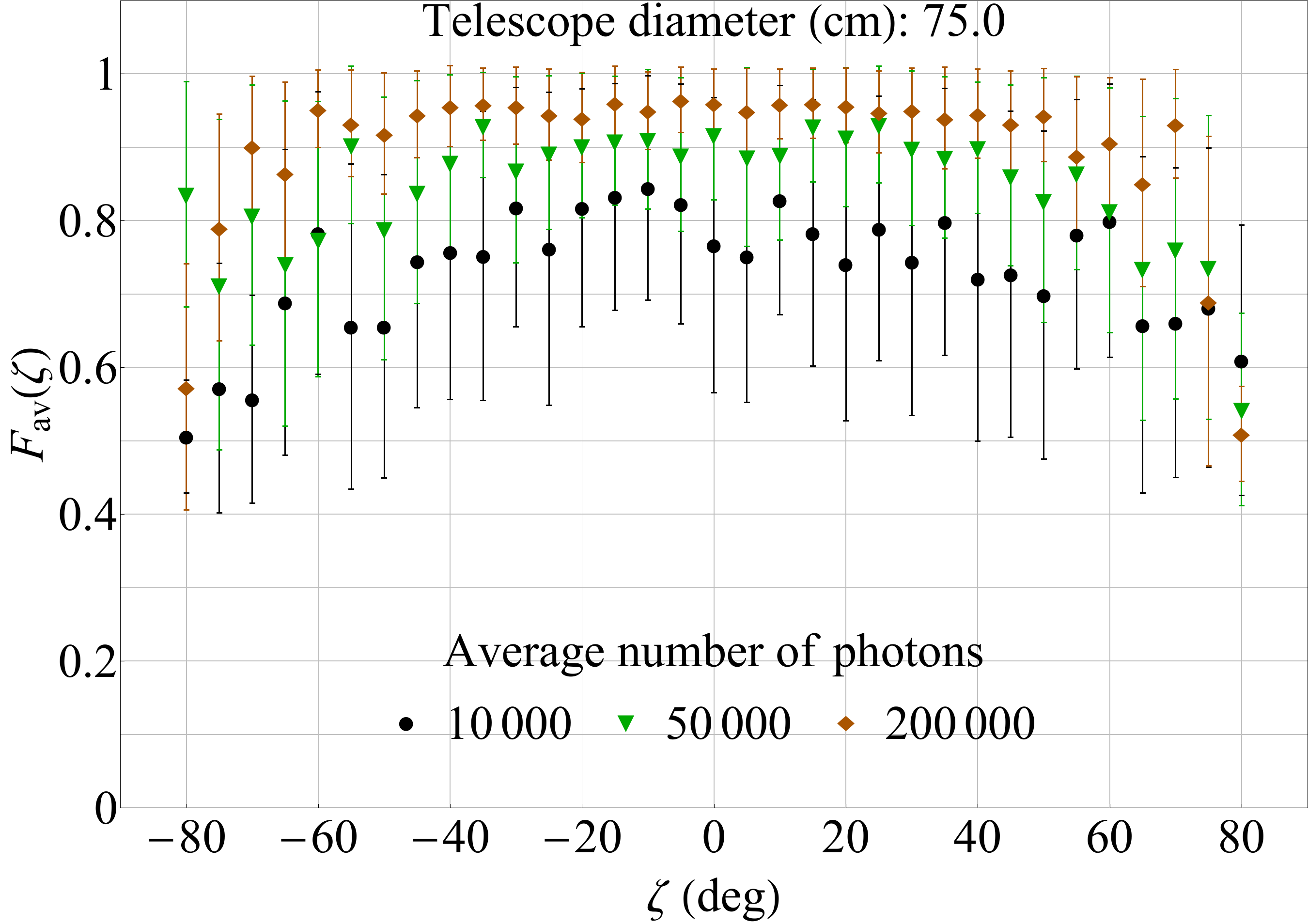}
        \put(-20,26){(c)}\\[2pt]
    \end{minipage}
    \hfill
    \begin{minipage}{0.48\textwidth}
        \centering
        \includegraphics[width=\linewidth]{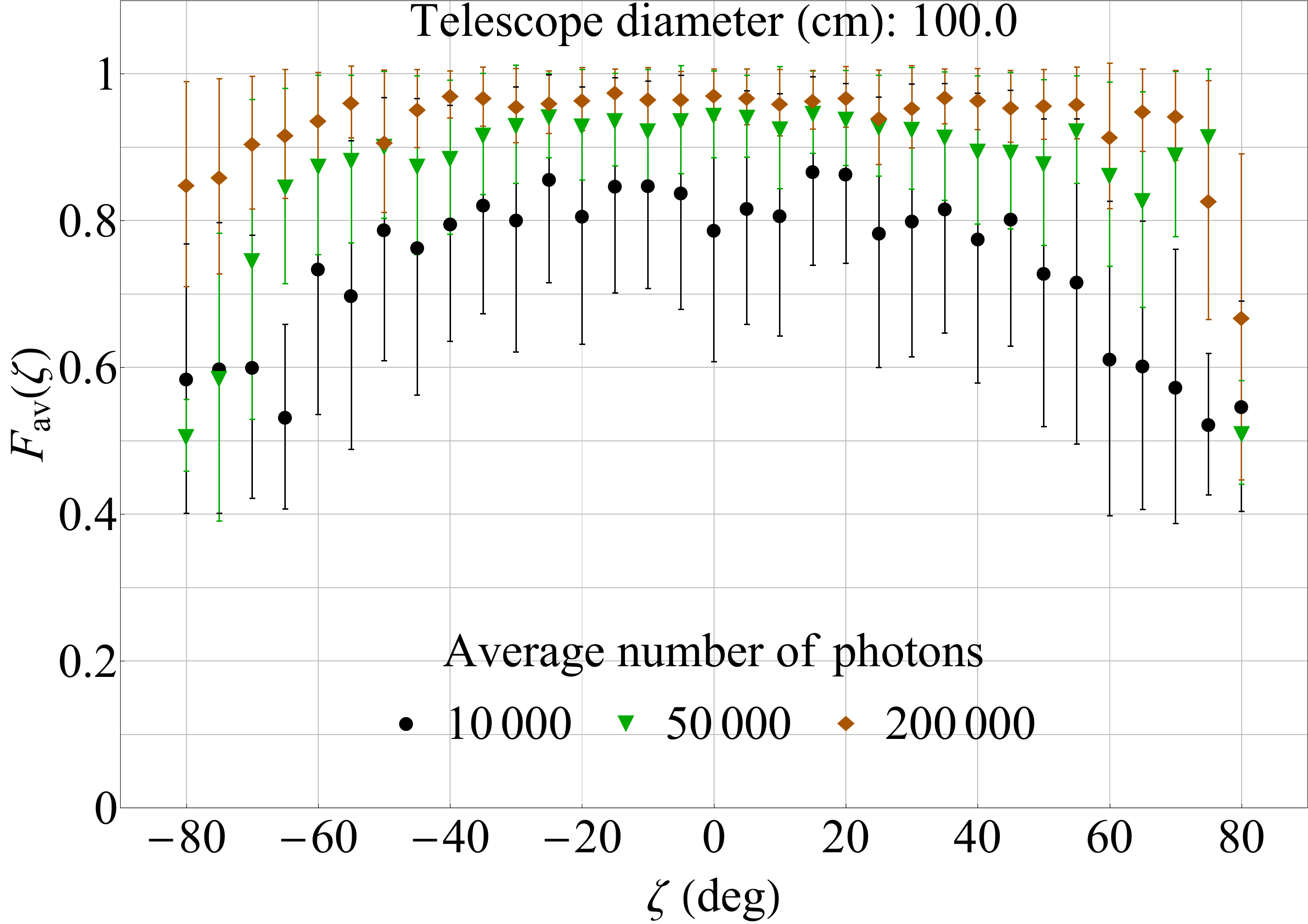}
        \put(-20,26){(d)}\\[2pt]
    \end{minipage}

    \caption{QST results for an LEO downlink (altitude of 420 km); each panel shows the reconstruction fidelity as a function of the zenith angle for four telescope diameters: (a) $D=25$ cm, (b) $D=50$ cm, (c) $D=75$ cm, and (d) $D=100$ cm.}
    \label{fig:set1}
\end{figure*}

\begin{figure*}[t]
    \centering

    \begin{minipage}{0.48\textwidth}
        \centering
        \includegraphics[width=\linewidth]
        {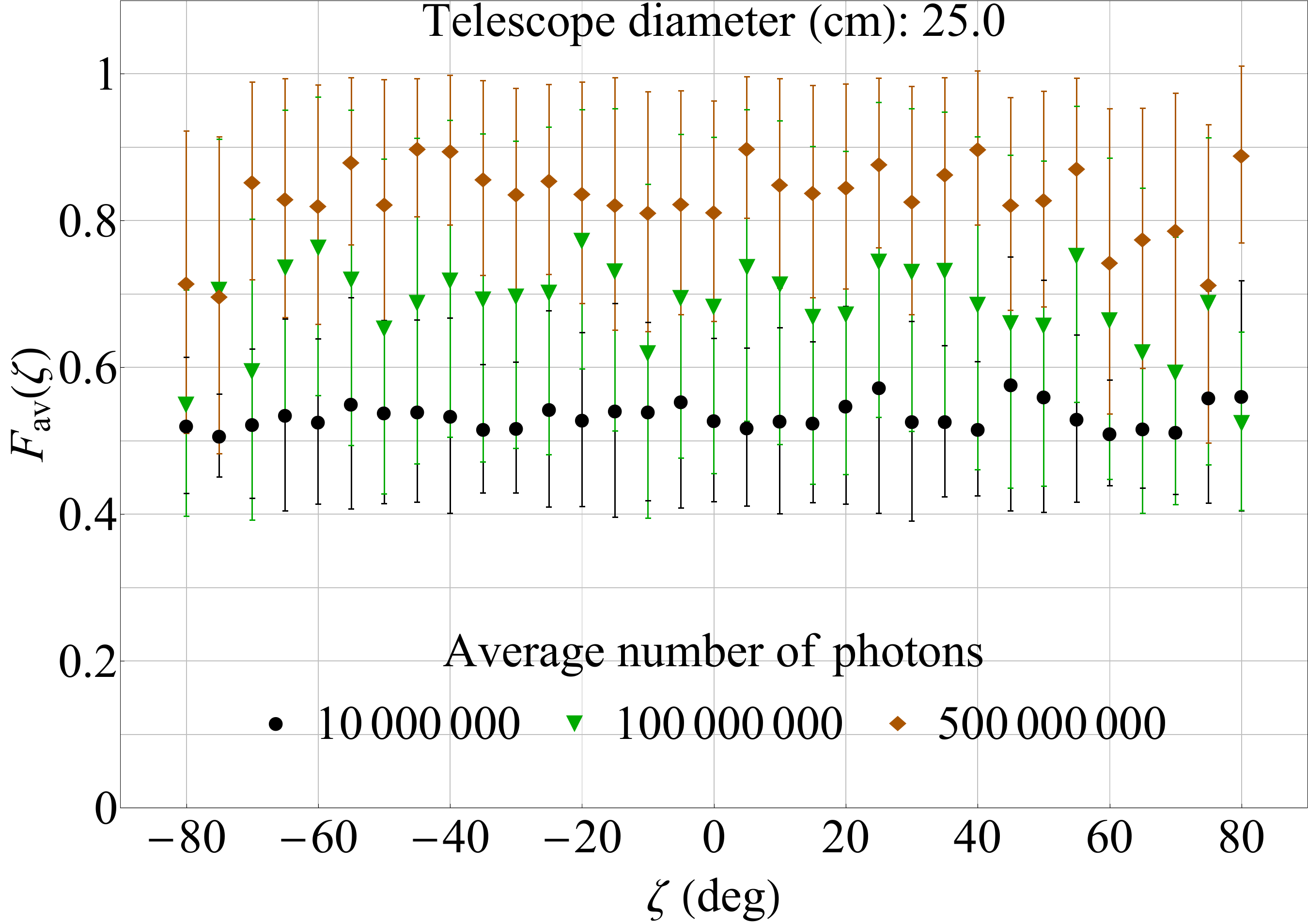}
        \put(-20,26){(a)}\\[2pt]
    \end{minipage}
    \hfill
    \begin{minipage}{0.48\textwidth}
        \centering
        \includegraphics[width=\linewidth]
        {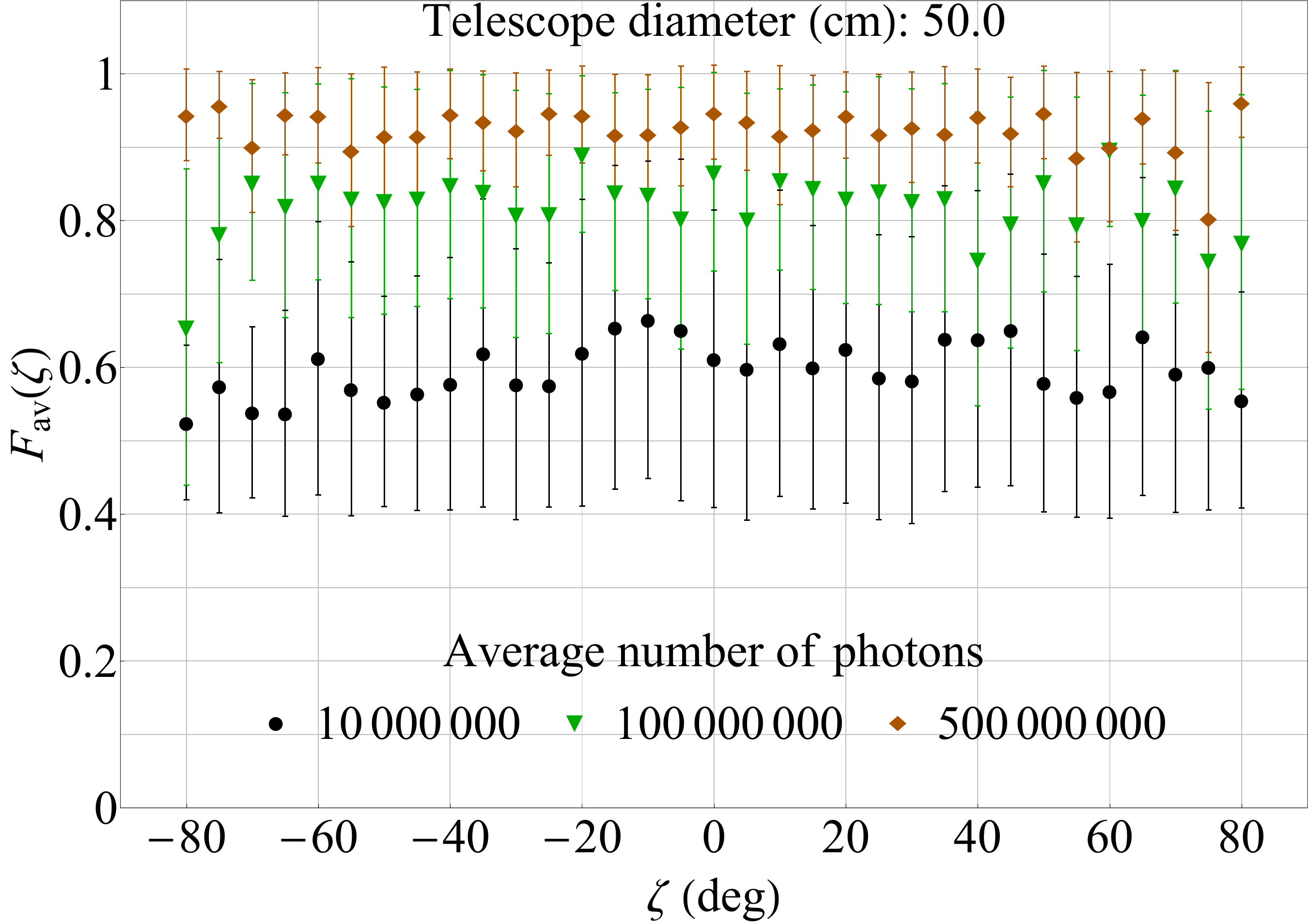}
        \put(-20,26){(b)}\\[2pt]
    \end{minipage}

    \vspace{1.8em}

    \begin{minipage}{0.48\textwidth}
        \centering
        \includegraphics[width=\linewidth]
        {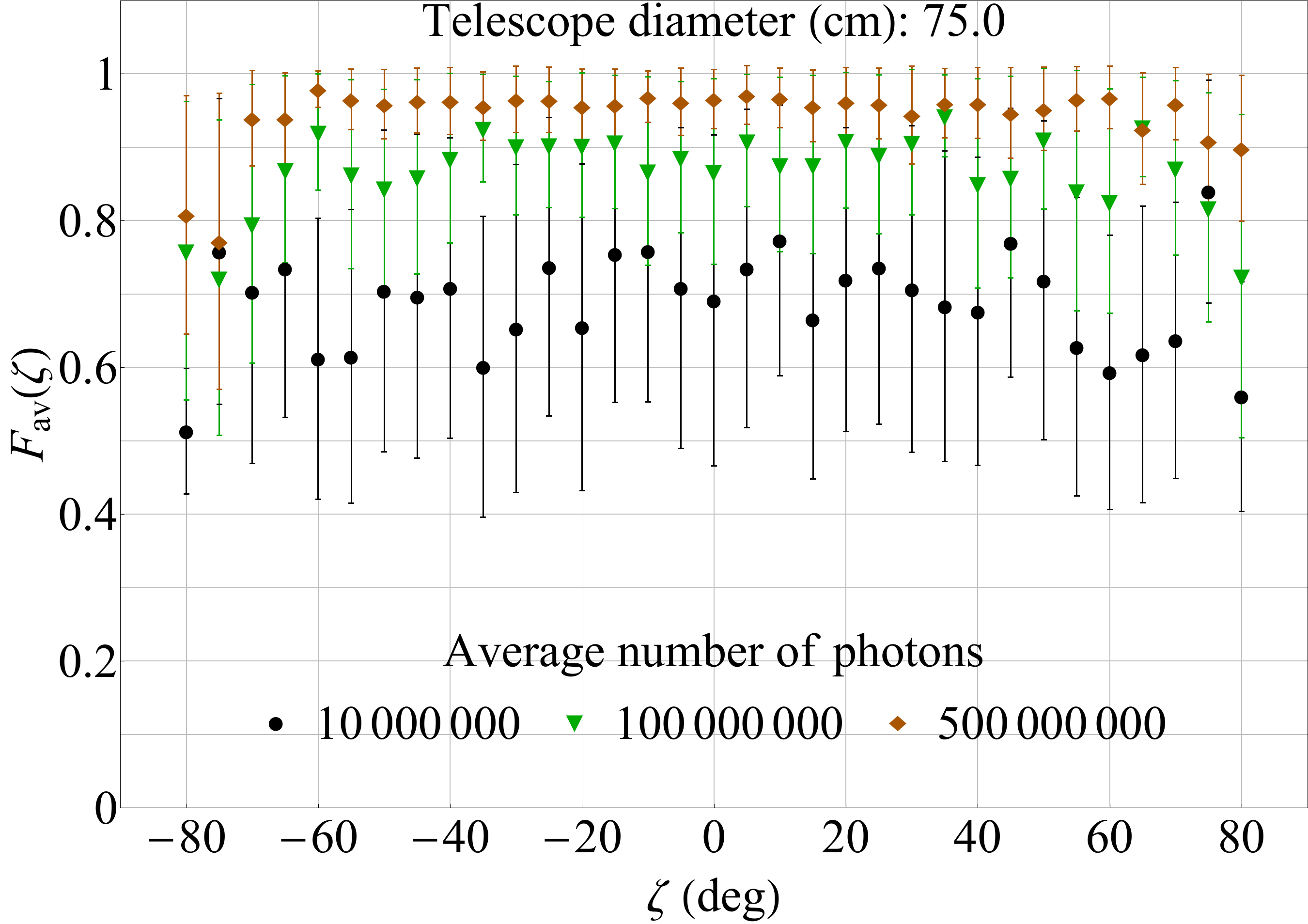}
        \put(-20,26){(c)}\\[2pt]
    \end{minipage}
    \hfill
    \begin{minipage}{0.48\textwidth}
        \centering
        \includegraphics[width=\linewidth]
        {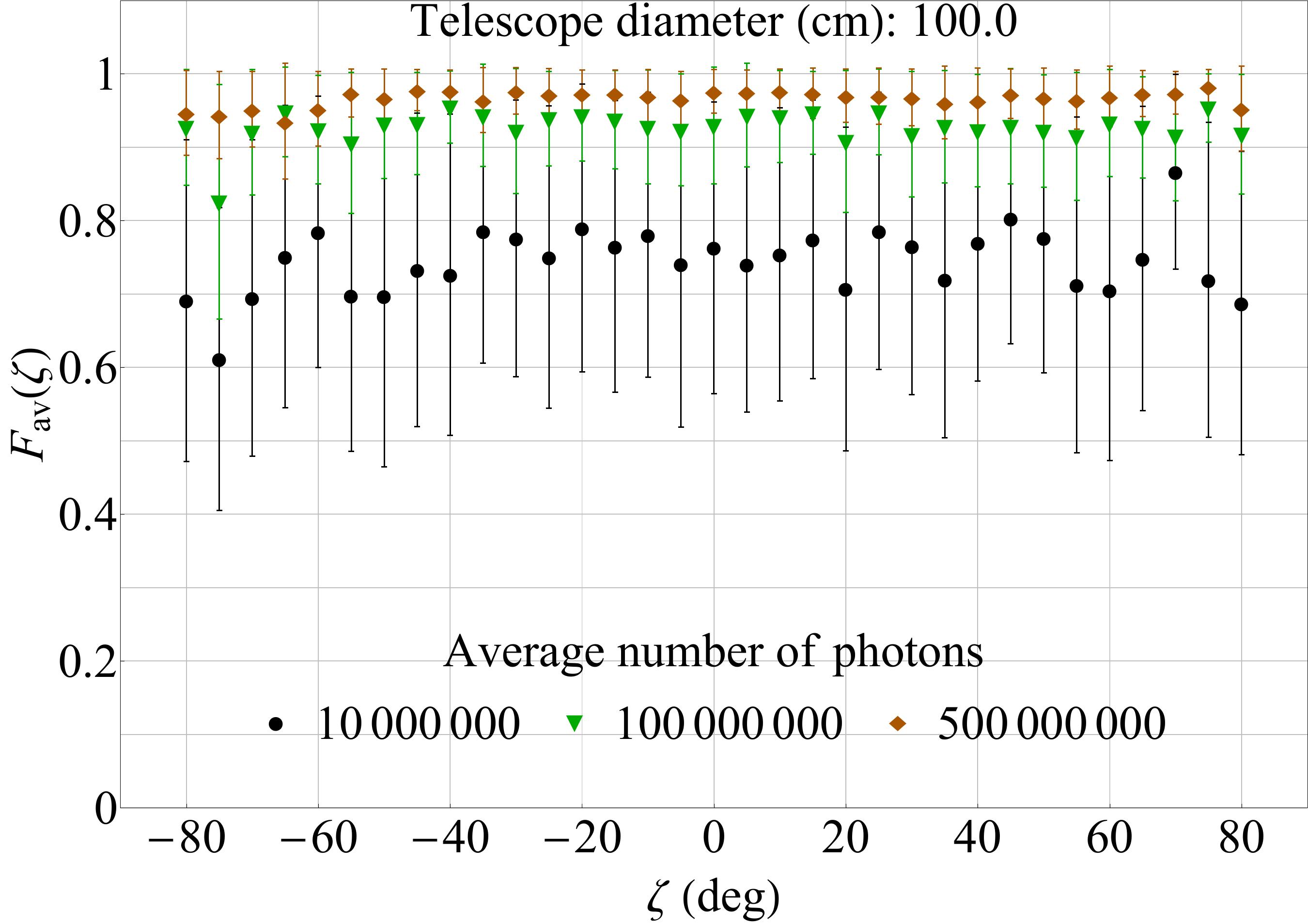}
        \put(-20,26){(d)}\\[2pt]
    \end{minipage}

    \caption{QST results for an MEO downlink (altitude of 20\,200 km): each panel shows the reconstruction fidelity as a function of the zenith angle for four telescope diameters: (a) $D=25$ cm, (b) $D=50$ cm, (c) $D=75$ cm, and (d) $D=100$ cm.}
    \label{fig:set2}
\end{figure*}

In this subsection, we address the practical question of how well one can reconstruct a quantum photonic state sent from LEO (Fig.~\ref{fig:set1}) and MEO (Fig.~\ref{fig:set2}) satellites as a function of the receiver aperture \(D\) and the mean number of photons \(\mathcal{N}\) emitted by the source for a single tomographic measurement setting. As explained in the previous subsection, \(\mathcal{N}\) represents the size of the ensemble of identically prepared photons used in the bulk tomography procedure. Different values of \(\mathcal{N}\) are considered across the subplots presented in Fig.~\ref{fig:set1} and Fig.~\ref{fig:set2}. The dominant physical effects determining performance are geometric collection (aperture area), atmospheric attenuation and turbulence (which grow with the zenith angle), and statistical error of the tomography (shot noise limited by the number of detected photons).

Figure~\ref{fig:set1} shows the average fidelity calculated from QST for qubits using different telescope sizes. The error bars correspond to one SD computed for the sample. These graphs refer to a LEO satellite at an altitude of $420$ km. The diameter of the telescope controls the collected signal power roughly as the receiving area $\propto D^2$. For fixed transmitted pulses, a larger aperture increases the detected photon number, which directly improves the signal-to-noise ratio (SNR) of the tomography. This is why the panels for $D=75$ cm and $D=100$ cm [Fig.~\ref{fig:set1}(c–d)] reach high fidelities with much smaller nominal photon budgets than the $D=25$ cm and $D=50$ cm panels [Fig.~\ref{fig:set1}(a–b)].

Moreover, each panel shows average fidelity as a function of zenith angle $\zeta$. At a larger $\zeta$, the optical path through the turbulent atmosphere is longer, producing larger extinction and stronger turbulence effects. Those effects reduce the detected photon rate and increase fluctuation-driven reconstruction errors, so fidelity degrades with increasing zenith angle for each curve.

The photon numbers used for the smaller apertures ($100$k–$1$M for $D=25/50$ cm) and for the larger apertures ($10$k–$200$k for $D=75/100$ cm) were chosen to illustrate the trade-off: a larger aperture can achieve the same fidelity with fewer photons. Fewer photons correspond to shorter signal durations since the average photon number is proportional to pulse duration under constant source brightness. As we know, shorter durations reduce exposure to slow, random atmospheric fluctuations during the measurement window and therefore can give better effective fidelity for the same nominal SNR when turbulence has significant low-frequency components. In practice, this means that for LEO downlinks, a larger receiver aperture allows one to achieve the same reconstruction fidelity with shorter acquisition times, thereby reducing the sensitivity of the tomography procedure to slow temporal fluctuations of the atmospheric channel.

Note that for typical LEO link losses (those presented in Fig.~\ref{fig:LEO_losses}(a)), tomography is feasible with modest photon budgets if the receiver aperture is $75–100$ cm, and even $25–50$ cm apertures can produce acceptable reconstructions provided the transmitted photon number is increased. The validity of these results is restricted to the set of system parameters listed in Table~\ref{tab1}, which were used to generate the simulated photon loss curves.

On the other hand, Fig.~\ref{fig:set2} examines the same tomography task for a MEO downlink. The key difference is the much larger geometric losses and consequently much lower channel transmittance compared to LEO. This imposes stronger quality requirements on the photon budget and/or aperture size for acceptable state reconstruction.
The MEO link has many times higher propagation losses than the LEO link used in Fig.~\ref{fig:set1}. Therefore, for a given photon budget, the number of photons reaching the ground is much smaller unless the receiver aperture is significantly increased. This is why the number of photons in Fig.~\ref{fig:set2} has risen to $10$ million, $100$ million, and $500$ million.

In the MEO case, there is a threshold behavior with respect to aperture size. For the small apertures shown ($25$ cm and $50$ cm), even the largest tested average number of photons (e.g., $10^7$) may still not have a sufficient number detected to produce useful tomography (this appears as the low-fidelity black curve). When the aperture is increased to 75 cm, the received photon flux enters a regime where tomography begins to produce meaningful fidelity, and $100$ cm further improves fidelity across zenith angles. This shows that for high-loss channels, enhancing the aperture is much more effective than moderate increases in emitted photons, because the aperture scales the collected flux as $D^2$, while increasing the emitted photons lengthens the acquisition and may still be subject to receiver and background limitations.

As in the LEO case, fidelity decreases with increasing zenith angle because of longer atmospheric paths and stronger turbulence.  However, for MEO, the dominant limiter near nadir is simply insufficient received photons. The effects of atmospheric turbulence become most important when the received flux is small, i.e. they amplify the poor SNR caused by long range.

For the simulation parameters used in Table~\ref{tab1} and the MEO losses in Fig.~\ref{fig:MEO_losses}(a), reliable QST from an MEO downlink requires either substantially larger receiver apertures ($\geq75$ cm in these plots) or very large photon budgets combined with careful background/detector control. Operationally, this pushes MEO tomography into a different engineering regime than LEO.

Compared to the LEO results in Fig.~\ref{fig:set1}, the MEO results in Fig.~\ref{fig:set2} operate in a fundamentally different loss regime, where the much larger propagation distance leads to orders-of-magnitude higher photon loss and therefore demands substantially larger photon numbers to achieve meaningful reconstruction fidelity. In the LEO case, fidelity improves smoothly with increasing telescope diameter, allowing a clear and flexible trade-off between receiver aperture and average photon number, whereas in the MEO case, this trade-off becomes strongly constrained and small apertures remain ineffective even at very high photon numbers. Although both figures show a degradation of fidelity with increasing zenith angle due to longer atmospheric paths, this effect is comparatively mild in LEO and becomes significantly more restrictive in MEO, where it compounds the already severe photon loss.

\subsection{Results of numerical simulations including background noise}

\begin{figure*}[t]
    \centering

    \begin{minipage}{0.48\textwidth}
        \centering
        \includegraphics[width=\linewidth]
        {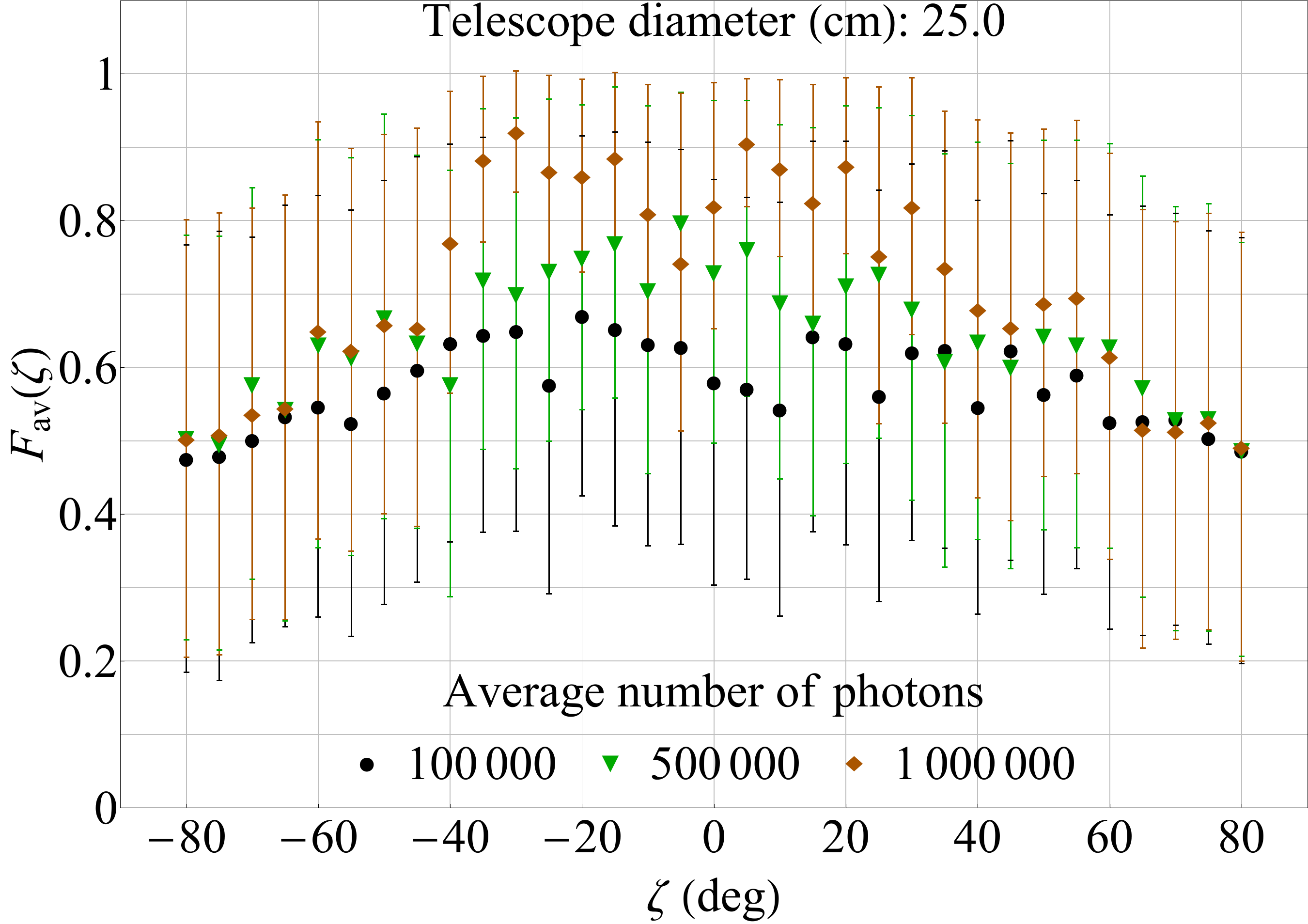}
        \put(-20,26){(a)}\\[2pt]
    \end{minipage}
    \hfill
    \begin{minipage}{0.48\textwidth}
        \centering
        \includegraphics[width=\linewidth]
        {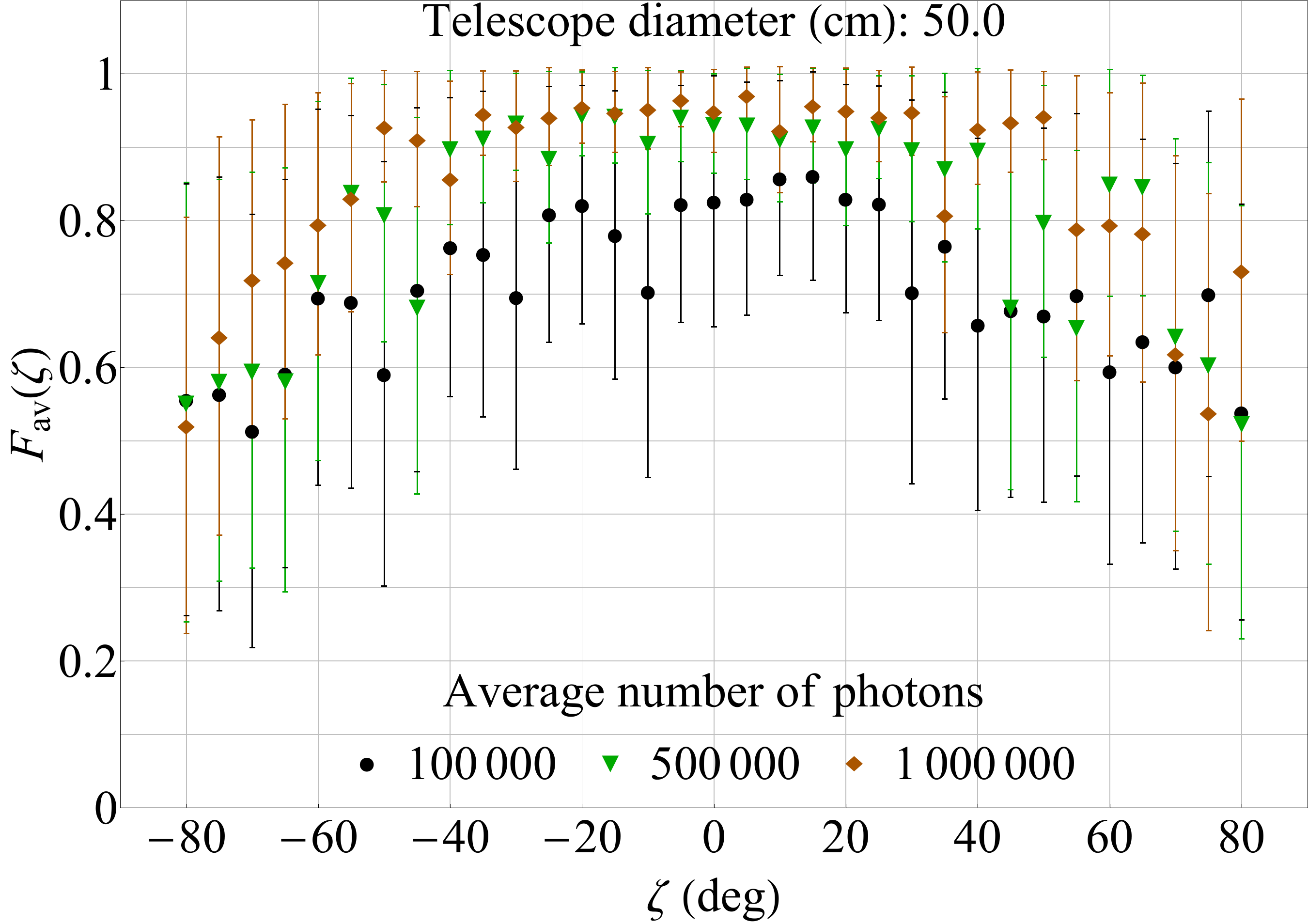}
        \put(-20,26){(b)}\\[2pt]
    \end{minipage}

    \vspace{1.8em}

    \begin{minipage}{0.48\textwidth}
        \centering
        \includegraphics[width=\linewidth]
        {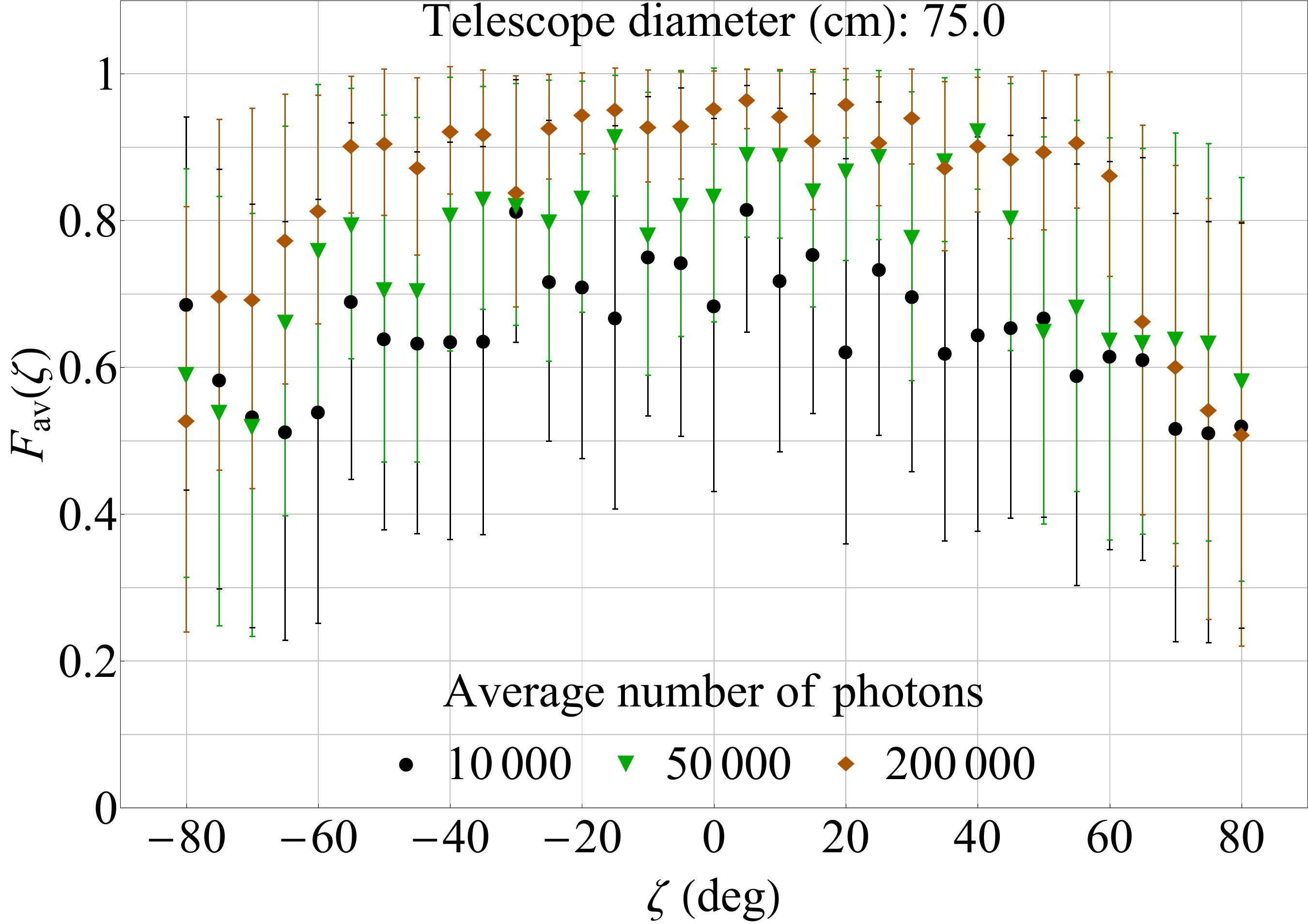}
        \put(-20,26){(c)}\\[2pt]
    \end{minipage}
    \hfill
    \begin{minipage}{0.48\textwidth}
        \centering
        \includegraphics[width=\linewidth]
        {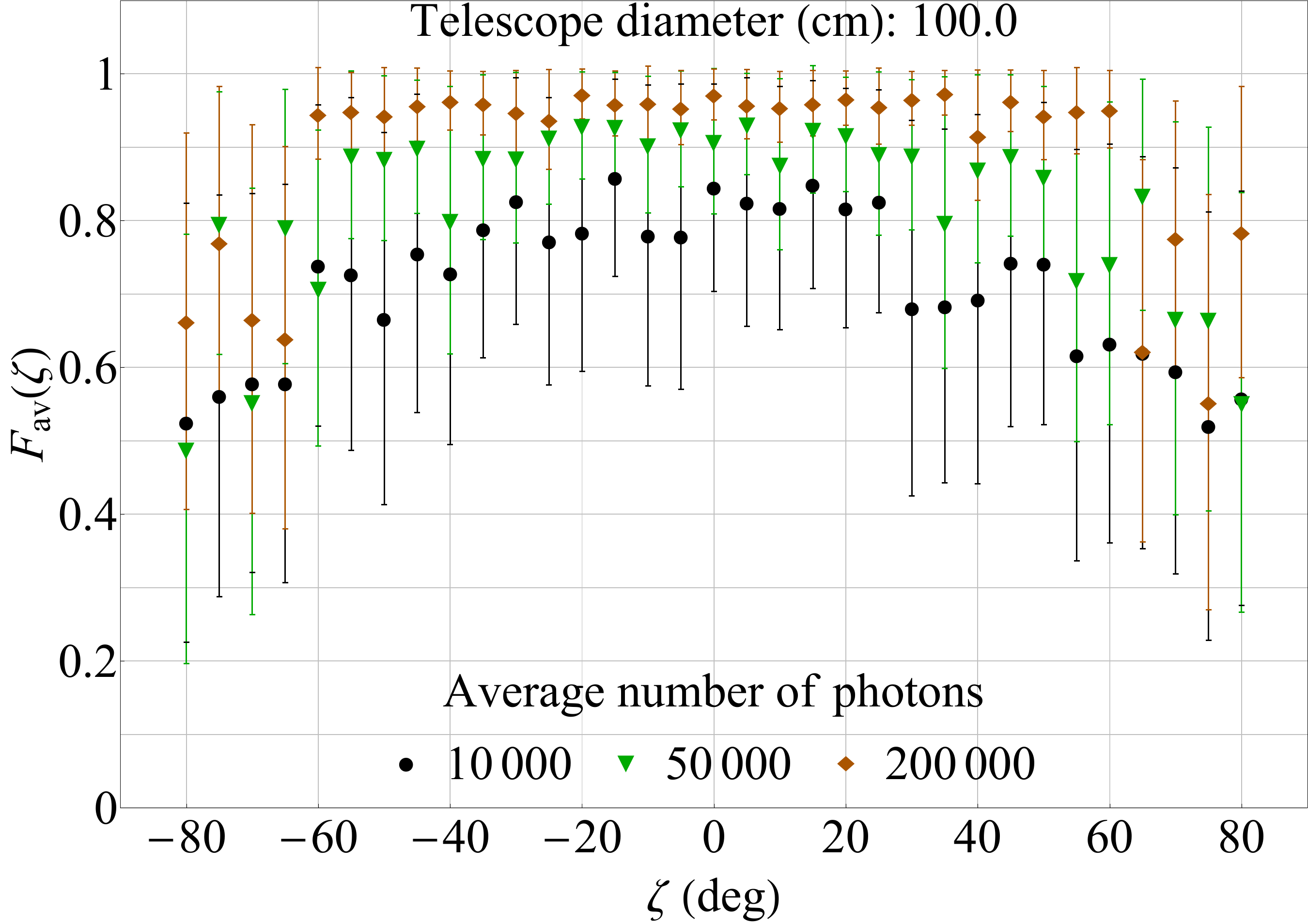}
        \put(-20,26){(d)}\\[2pt]
    \end{minipage}

    \caption{QST results for an LEO downlink (altitude of 420 km) with background counts: each panel shows the reconstruction fidelity as a function of the zenith angle for four telescope diameters: (a) $D=25$ cm, (b) $D=50$ cm, (c) $D=75$ cm, and (d) $D=100$ cm. Sky brightness $B_\lambda = 10^{-3} \,\mathrm{W\,m^{-2}\,nm^{-1}\,sr^{-1}}$.}
    \label{fig:setLEObright}
\end{figure*}

The formula Eq.~(\ref{measuredcounts}) for the simulation of the registered photon counts does not involve any excess background noise. However, in realistic free-space optical systems, the performance of QST is affected not only by atmospheric attenuation and turbulence, but also by background photons collected by the receiver. Such photons originate primarily from sky radiance and may contribute to accidental detection events, thereby reducing the fidelity of the reconstructed quantum state. In order to estimate this effect, we extend our QST framework by including a model for background photon counts collected during the tomographic acquisition window.

The expected number of background photons registered by the detection system is modeled as \cite{pirandola2021limits}
\begin{equation}
    N_{\mathrm{bg}}
    =
    \frac{\lambda}{h c}\,
    B_\lambda\,
    \tau\,
    \Delta\lambda\,
    \Omega_{\mathrm{FOV}}\,
    A_{\mathrm{tel}}\,
    \eta_{\mathrm{int}},
\end{equation}
where $\lambda$ denotes the optical wavelength, $h$ is the Planck constant, and $c$ is the speed of light in vacuum. The quantity $B_\lambda$ represents the spectral sky radiance expressed (in units of \(\mathrm{W\,m^{-2}\,nm^{-1}\,sr^{-1}}\) and called brightness for short) while $\tau$ is the effective acquisition time (detection window), $\Delta\lambda$ is the bandwidth of the optical filter, $\Omega_{\mathrm{FOV}}$ is the field of view of the receiver expressed in steradians, $A_{\mathrm{tel}}$ is the telescope aperture area, and $\eta_{\mathrm{int}}$ denotes the overall internal detection efficiency.

For the numerical simulations presented in this section, we assume \(\lambda\) and \(\eta_{\mathrm{int}} \) to take the same values as in Tab.~\ref{tab1}, while an optical filter bandwidth is \(\Delta\lambda = 1~\mathrm{nm}\). Moreover, we adopt two representative sky radiance values
\begin{equation}
B_\lambda = 10^{-3} \hspace{0.25cm}\text{or}\hspace{0.25cm} 10^{-5} \;
(\mathrm{W\,m^{-2}\,nm^{-1}\,sr^{-1}}),
\end{equation}
which approximately correspond to bright civil twilight conditions and darker astronomical twilight conditions, respectively. Such values represent scenarios with substantial background illumination caused by scattered sunlight, even though the Sun is already below the horizon.

The acquisition time $\tau$ is directly related to the number of photons used in the tomography procedure. In our model, we assume that the average photon number emitted by the source is proportional to the pulse duration for a fixed source brightness. Consequently, larger photon ensembles correspond to longer acquisition windows. To ensure physical consistency with atmospheric dynamics, the acquisition window is selected such that even the largest investigated photon ensemble remains within the characteristic atmospheric coherence time, which is typically of the order of milliseconds. More specifically, the scaling is chosen so that the largest investigated photon number, corresponding to $5\times10^8$ photons, produces an acquisition window of approximately $1$ ms. Since a complete SIC-POVM tomography round for a qubit requires four measurement settings, the total duration of one tomographic reconstruction remains approximately $4$ ms, which is consistent with the assumption of quasi-stationary atmospheric conditions during the measurement process.

Then, for each SIC-POVM measurement operator, we simulate the number of background photons reaching the receiver by sampling from a Poisson distribution,
\[
m^{\mathrm{bg}}_k \sim \mathrm{Pois}(N_{\mathrm{bg}}).
\]
Finally, under bright-sky conditions, the registered photon counts are modeled as
\begin{equation}
    m_k^{\mathrm{bright}}= m_k +\frac{1}{2} m^{\mathrm{bg}}_k,
\end{equation}
where $m_k$ is computed according to Eq.~(\ref{measuredcounts}). The factor $1/2$ appears because the background photons are assumed to be in a maximally mixed polarization state. Consequently, only half of them contribute on average to clicks associated with a given polarization measurement outcome.

\begin{figure*}[t]
    \centering

    \begin{minipage}{0.48\textwidth}
        \centering
        \includegraphics[width=\linewidth]
        {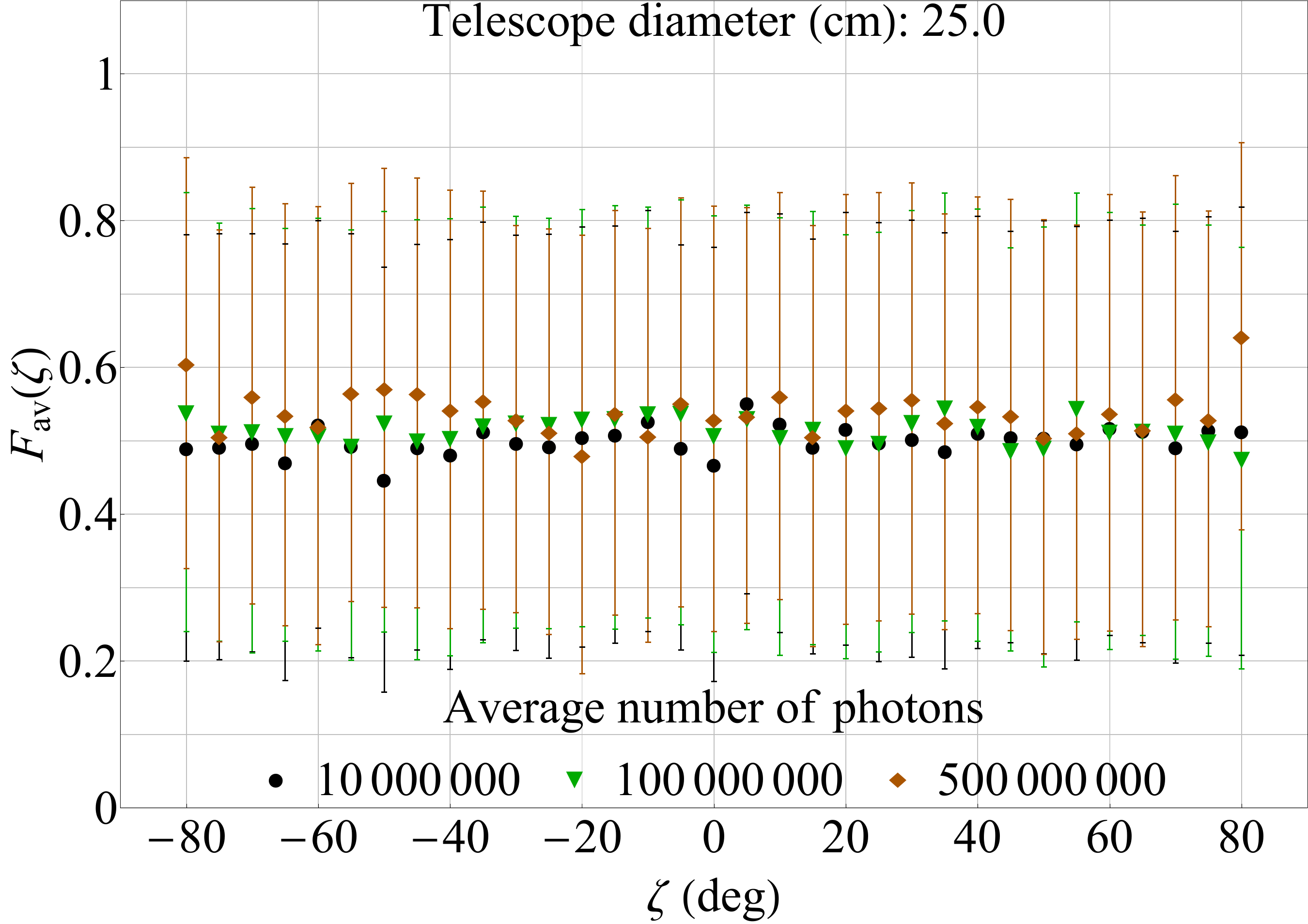}
        \put(-20,26){(a)}\\[2pt]
    \end{minipage}
    \hfill
    \begin{minipage}{0.48\textwidth}
        \centering
        \includegraphics[width=\linewidth]
        {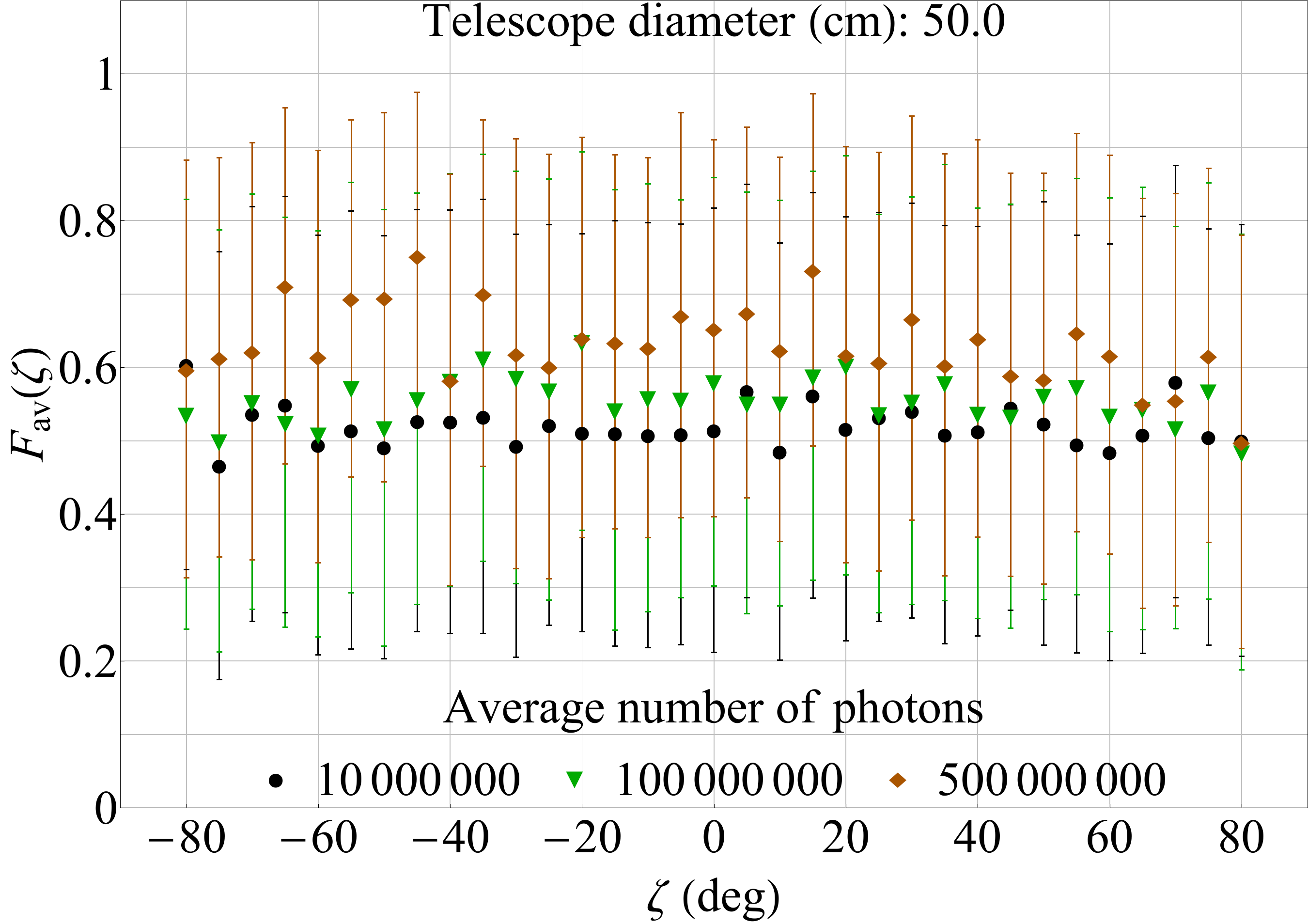}
        \put(-20,26){(b)}\\[2pt]
    \end{minipage}

    \vspace{1.8em}

    \begin{minipage}{0.48\textwidth}
        \centering
        \includegraphics[width=\linewidth]
        {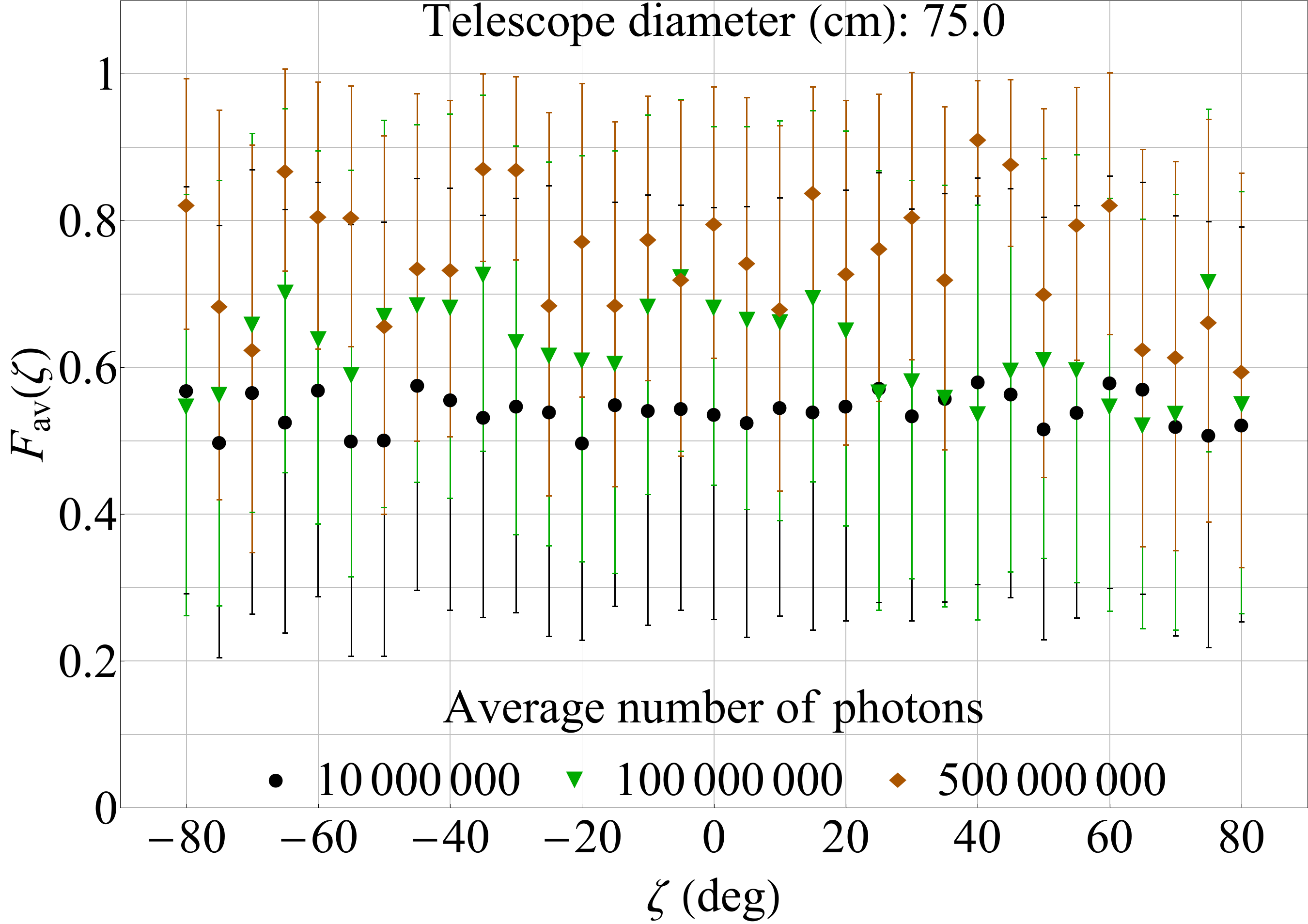}
        \put(-20,26){(c)}\\[2pt]
    \end{minipage}
    \hfill
    \begin{minipage}{0.48\textwidth}
        \centering
        \includegraphics[width=\linewidth]
        {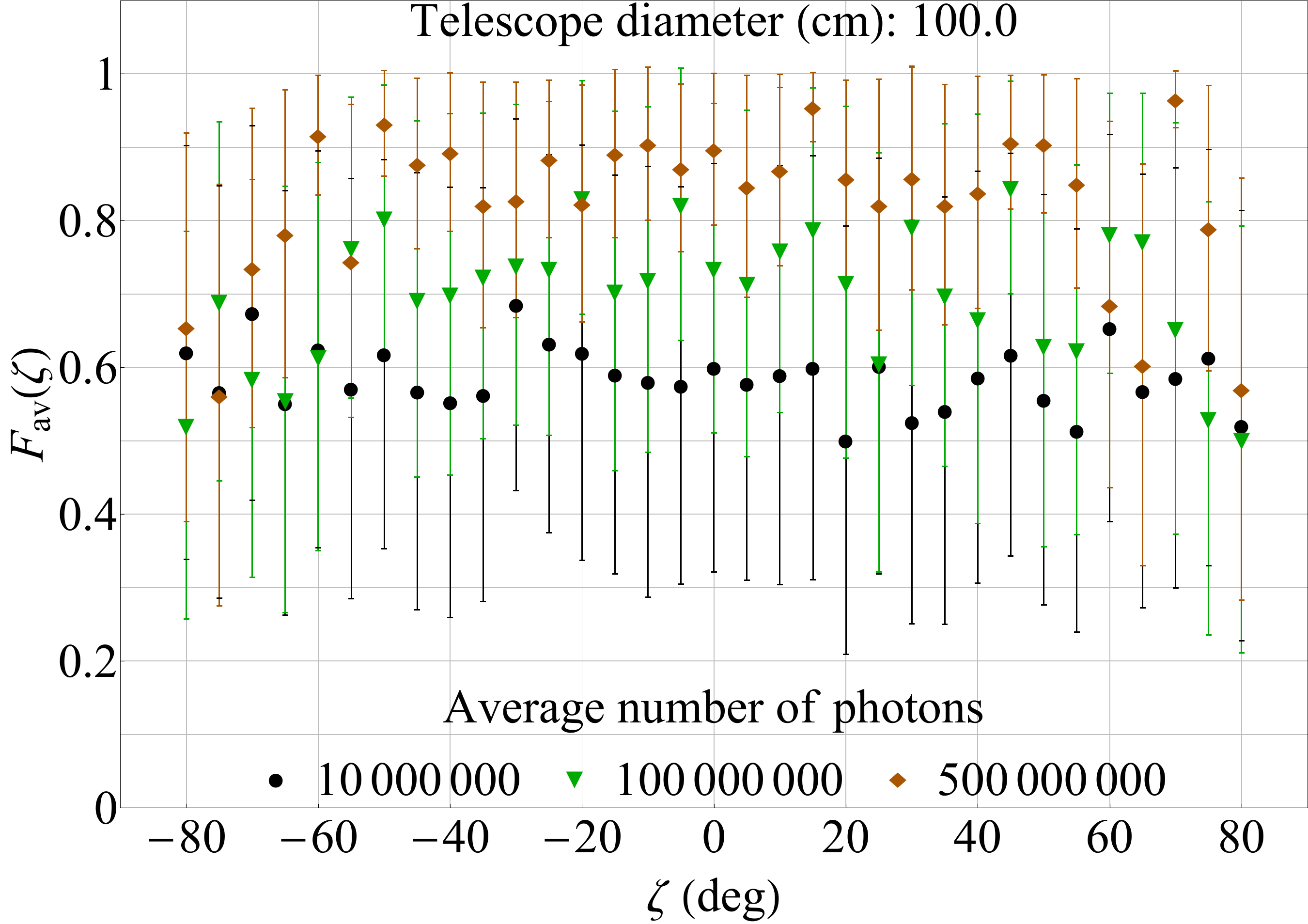}
        \put(-20,26){(d)}\\[2pt]
    \end{minipage}

    \caption{QST results for an MEO downlink (altitude of 20\,200 km) with background counts: each panel shows the reconstruction fidelity as a function of the zenith angle for four telescope diameters: (a) $D=25$ cm, (b) $D=50$ cm, (c) $D=75$ cm, and (d) $D=100$ cm. $B_\lambda = 10^{-5} \,\mathrm{W\,m^{-2}\,nm^{-1}\,sr^{-1}}$.}
    \label{fig:setMEObright}
\end{figure*}

The results that include background counts are presented in Fig.~\ref{fig:setLEObright} for a LEO satellite and in Fig.~\ref{fig:setMEObright} for a MEO satellite. The corresponding sky brightness values are assumed to be \(B_\lambda = 10^{-3}\) and \(B_\lambda = 10^{-5}\,\mathrm{W\,m^{-2}\,nm^{-1}\,sr^{-1}}\), respectively. The lower value adopted for the MEO case was chosen deliberately, because higher sky brightness values led to a clear degradation of the reconstruction fidelity. This results from the large propagation losses in the MEO downlink and the need to use larger photon budgets, which, for a fixed source brightness, correspond to longer acquisition windows. Longer detection windows increase the number of registered background photons. Therefore, the MEO case was evaluated under darker-sky conditions in order to present an operational regime in which the background contribution does not completely dominate the QST reconstruction process.

We first consider the LEO case with background counts originating from sky brightness (Fig.~\ref{fig:setLEObright}), where the average QST reconstruction fidelity is plotted for different telescope diameters at an altitude of \(420~\mathrm{km}\). As the telescope aperture increases for a fixed photon budget, the reconstruction fidelity generally improves, because a larger receiver collects more signal photons. A similar effect is observed when the mean number of photons is increased: curves corresponding to larger photon budgets reach higher values of the average fidelity. The fidelity also depends on the zenith angle, the best results are obtained mainly in the central range of angles, whereas for large values of \(|\zeta|\) the fidelity decreases because the longer propagation path through the atmosphere increases attenuation and turbulence effects. Under bright-sky conditions, additional background counts can further reduce the effective signal-to-noise ratio, especially at larger zenith angles, where the number of received signal photons is smaller.

A comparison of Fig.~\ref{fig:set1} and Fig.~\ref{fig:setLEObright} shows that including background counts leads to a degradation of the reconstruction fidelity and an increase in the statistical spread of the results, as reflected by larger and more irregular error bars. Additional photons originating from sky brightness generate accidental detection events that do not carry information about the reconstructed quantum state and therefore reduce the effective signal-to-noise ratio. This effect is particularly visible for smaller apertures, \(D = 25~\mathrm{cm}\) and \(D = 50~\mathrm{cm}\), especially for lower photon budgets considered within each aperture case, as well as for large zenith angles. For larger apertures, \(D = 75~\mathrm{cm}\) and \(D = 100~\mathrm{cm}\), and for higher photon budgets considered within these cases, the influence of the background is less significant, because the number of registered signal photons is larger and partially dominates over the accidental background counts. However, even in these cases, the degradation remains noticeable for large zenith angles.

The QST reconstruction results for the MEO downlink at an altitude of \(20\,200~\mathrm{km}\), including background counts originating from sky brightness, are presented in Fig.~\ref{fig:setMEObright}. It can be seen that for smaller apertures, \(D = 25~\mathrm{cm}\) and \(D = 50~\mathrm{cm}\), the reconstruction fidelity remains relatively low and is affected by large uncertainties, even for very large photon budgets. For larger apertures, \(D = 75~\mathrm{cm}\) and \(D = 100~\mathrm{cm}\), the values of average fidelity increase noticeably, especially for the largest number of photons, indicating that increasing both the aperture and the photon budget partially compensates for the large propagation losses characteristic of the MEO downlink and the adverse effect of background counts.

A comparison between Fig.~\ref{fig:set2} and Fig.~\ref{fig:setMEObright} shows that the inclusion of background counts in the MEO link leads to a clear reduction in the average QST reconstruction fidelity and an increase in the statistical spread of the obtained results, especially for smaller receiver apertures, \(D = 25~\mathrm{cm}\) and \(D = 50~\mathrm{cm}\). Additional photons originating from sky brightness increase the number of accidental detection events, which reduces the effective signal-to-noise ratio and results in larger uncertainties in the average fidelity values. This effect is particularly important because the MEO link is characterized by large propagation losses, so the number of signal photons reaching the receiver is limited. For larger apertures, \(D = 75~\mathrm{cm}\) and \(D = 100~\mathrm{cm}\), and for the largest photon budget, the influence of the background is partially compensated. However, the average fidelity values remain lower and larger statistical spread than in the case without background counts.

\vspace{1cm}
\section{Conclusions and Outlook}\label{finalsec}

In this paper, we have analyzed the link budget for optical downlink communication from LEO and MEO satellites. Estimating the link budget is a crucial step in the planning of satellite missions, as it allows one to determine the required optical power of the source to ensure that the received signal at the OGS meets specific performance criteria. Importantly, link budget analysis, which quantifies photon loss over an FSO link, is equally relevant in classical scenarios (e.g., optical ranging or optical key distribution \cite{ref2,Czerwinski2023,Czerwinski2026}) and quantum applications (e.g., satellite QKD or entanglement distribution).

As part of this analysis, we incorporated the effects of aperture averaging on the fluctuations in transmittance caused by atmospheric turbulence. Using the model proposed by Andrews and Phillips~\cite{andrews2005laser}, we evaluated how increasing the receiver aperture diameter can suppress turbulence-induced intensity fluctuations, and how this suppression affects the total photon loss for different zenith angles.

Several key conclusions can be drawn from our results. First, aperture averaging proves to be effective for LEO downlinks. At typical LEO altitudes, even medium-size telescope diameters (e.g., 0.25~m) significantly mitigate scintillation effects. This leads to more stable received signal power and reduced variability in photon counts, which is particularly beneficial for quantum applications that rely on consistent link quality.

For MEO downlinks, the effectiveness of aperture averaging is more limited due to the much greater propagation distance. The same aperture sizes that perform well for LEO links offer only marginal benefits for MEO links. Although increasing the receiver diameter does enhance signal collection, its influence on reducing intensity fluctuations is small and primarily noticeable at low zenith angles.

These findings emphasize the role of telescope size as a critical design parameter. In LEO missions, the aperture can be optimized to balance cost, mass, and performance. For MEO links, however, larger apertures mainly serve to improve signal strength, with limited impact on turbulence mitigation. To effectively mitigate turbulence-induced fluctuations in higher orbits, we need other complementary approaches, such as adaptive optics or advanced signal processing.

QST in satellite downlinks demonstrates a clear dependence on orbital altitude through its effect on channel loss and photon collection efficiency. In LEO scenarios, the comparatively moderate losses allow high-fidelity state reconstruction utilizing realistic photon numbers, with a flexible trade-off between receiver aperture and acquisition time that enables robust and efficient tomography. For MEO links, the much higher attenuation fundamentally shifts the operating regime, making large receiver apertures and very high photon budgets essential for obtaining statistically meaningful reconstructions. In this high-loss regime, small apertures remain ineffective even for long measurement times, further amplifying atmospheric effects on reconstruction errors. These results show that QST is feasible for both LEO and MEO platforms, provided that the system parameters are carefully tuned to suit the link-loss regime, and confirm its suitability as a practical tool for in-orbit quantum state verification.

Including background counts demonstrates that sky brightness can significantly affect the fidelity of quantum state tomography, particularly when only a small number of signal photons reach the receiver. The impact of background noise can be mitigated by increasing the receiver aperture and employing larger photon budgets. However, these improvements involve nontrivial trade-offs. Increasing the photon budget generally requires longer acquisition times, which also increase the number of accumulated background counts. Similarly, a larger receiver aperture enhances channel transmittance and increases the number of detected signal photons, but it also collects more background radiation. Consequently, the relationship between reconstruction fidelity and these system parameters is not characterized by a simple linear dependence. Nevertheless, our results indicate that background counts do not prevent successful state reconstruction under realistic operating conditions, although they remain an important factor limiting the achievable QST performance.

A limitation of the present study is that the transmittance model does not include pointing, acquisition, and tracking (PAT) errors or beam wandering effects. Throughout this work, ideal beam alignment between the satellite terminal and the optical ground station is assumed. So, the reported transmittance values account only for atmospheric extinction, diffraction-induced losses, and turbulence-induced intensity fluctuations. In practical systems, residual pointing inaccuracies can introduce additional attenuation, particularly for long-distance links and narrow beam divergences. These effects should therefore be incorporated into future end-to-end performance analyses.

Another source of discrepancy between the proposed model and practical implementations may arise from internal losses within the optical terminals. In our framework, such effects are represented by the factor $\eta_{\mathrm{int}}$. However, since this study does not assume a specific hardware implementation, detector technology, or optical architecture, it is not possible to estimate this parameter more precisely. In real systems, internal losses depend on numerous factors, including optical component efficiencies, coupling losses, spectral filtering, detector performance, and system integration. As a result, they may contribute significantly to the overall link budget and further increase the total photon loss beyond the values predicted by the present model.

Several directions for future research arise from this study. It would be natural to extend the analysis to uplink scenarios, where the beam undergoes spreading early in its path, making it more susceptible to turbulence. Such studies might reveal different dependencies on aperture size. Furthermore, simulations that include the temporal dynamics of scintillation and outage probability under varying atmospheric conditions could provide deeper insights into link reliability. Finally, future work should aim to incorporate these effects into quantum link budget models, accounting not only for intensity loss but also for phase distortions and entanglement degradation. These efforts will support the development of robust and scalable FSO systems for both classical and quantum communication networks.

\section*{Funding}
J.J.B. was supported by the National Science Centre of Poland (Grant No. 2020/39/O/ST2/00137).

\section*{Data Availability}
The data that support the findings of this article are publicly available from Ref. \cite{CzerwinskiBorkowskiData}

\section*{Conflict of Interest}
The author declares that there is no conflict of interest.

\end{document}